\documentclass{jfm}
\usepackage{graphicx}
\usepackage{epstopdf, epsfig, color, xcolor}
\usepackage{subfigure}
\usepackage[T1]{fontenc}
\usepackage{newtxtext}
\usepackage[varvw]{newtxmath}

\newcommand{\pard}[2]{\frac{\partial #1}{\partial #2}}

\shorttitle{Onset of global instability in a premixed annular V-flame}
\shortauthor{C.-H. Wang, C. Douglas, Y. Guan, C.-X. Xu and L. Lesshafft}

\title{Onset of global instability in a premixed annular V-flame}

\author{Chuhan Wang\aff{1,2}\corresp{\email{cwang@ladhyx.polytechnique.fr}}, Christopher M. Douglas\aff{2,3}, Yu Guan\aff{4}, Chunxiao Xu\aff{1} and Lutz Lesshafft\aff{2}}

\affiliation{
\aff{1}AML, Department of Engineering Mechanics, Tsinghua University, 100084 Beijing, PR China
\aff{2}LadHyX, CNRS, École Polytechnique, Institut Polytechnique de Paris, 91120 Palaiseau, France 
\aff{3}Department of Mechanical Engineering and Materials Science, Duke University, Durham, NC, USA
\aff{4}Department of Aeronautical and Aviation Engineering, The Hong Kong Polytechnic University, Kowloon, Hong Kong}

\begin{document}

\maketitle

\begin{abstract}
	We investigate self-excited axisymmetric oscillations of a lean premixed methane--air V-flame in a laminar annular jet. The flame is anchored near the rim of the centrebody, forming an inverted cone, while the strongest vorticity is concentrated along the outer shear layer of the annular jet. Consequently, the reaction and vorticity dynamics are largely separated, except where they coalesce near the flame tip. The global eigenmodes corresponding to the linearised reacting flow equations around the steady base state are computed in an axisymmetric setting. We identify an arc branch of eigenmodes exhibiting strong oscillations at the flame tip. The associated eigenvalues are robust with respect to domain truncation and numerical discretisation, and they become destabilised as the Reynolds number increases. The frequency of the leading eigenmode is found to correspond to the Lagrangian disturbance advection time from the nozzle outlet to the flame tip. The essential role of this convective mechanism is also supported by resolvent analysis, which finds that the same flame-tip disturbance structure and frequency are optimally amplified when the flame is subjected to external white noise forcing. Strong non-modal effects in the form of pseudo-resonance are not found. Nonlinear time-resolved simulation further reveals notable hysteresis phenomena in the subcritical regime prior to instability. Hence, even when the flame is linearly stable, perturbations of sufficient amplitude can trigger limit-cycle oscillations and higher-dimensional dynamics sustained by nonlinear feedback. A Monte Carlo simulation of passive tracers in the unsteady flame suggests a nonlinear non-local instability mechanism. Notably, linear analysis of the subcritical time-averaged limit-cycle state yields eigenvalues that do not match the nonlinear periodic oscillation frequencies. This mismatch is attributed to the fundamentally nonlinear dynamics of the subcritical V-flame instability, where the dichromatic, non-local interaction between the heat release rate along the flame surface and the vortex dynamics in the jet shear layer cannot be approximated as a simple distortion of the mean flow.
\end{abstract}

\begin{keywords}
\end{keywords}

\section{Introduction}
\begin{figure}
	\centerline{\includegraphics[width=\textwidth]{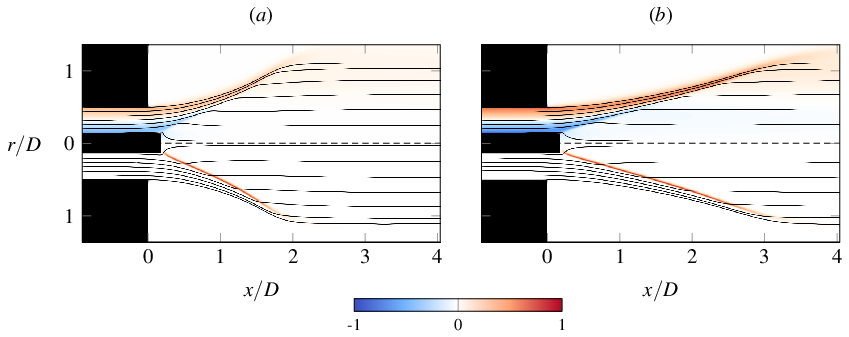}}
	\caption{Steady base states of the calculated V-flame at (\textit{a}) $\Rey=1674$  and (\textit{b}) $\Rey=2891$. Top of each plot: azimuthal vorticity $\Omega_z$. Bottom of each plot: heat release rate $\dot{\omega}_T$. Both variables are normalised by their maximum. Streamlines are superposed.}
	\label{fig:base}
\end{figure}

Canonical configurations of premixed laminar flames attract significant research interest because their unsteady dynamics are rich and representative of many practical configurations, yet relatively simple in comparison to turbulent flames \citep{lieuwen2003modeling, schuller2020dynamics}. A V-flame, the configuration considered in the present study, is an inverted conical flame anchored on the centrebody of an annular burner. In industrial applications, V-flames are often enhanced with swirl to stabilise the flame and prevent blow-off \citep{candel2014dynamics}. However, even in the absence of swirl, V-flame dynamics remain intricate. Flow fields corresponding to a non-swirling V-flame, computed on a two-dimensional axisymmetric grid at two different Reynolds numbers, are presented in figure~\ref{fig:base}. The flame surface region, characterised by high-magnitude heat release rates, is anchored to the centrebody. A strong free shear layer is shed from the outer corner of the jet nozzle, as illustrated by the vorticity distribution. Unlike the cylinder-anchored flame investigated in our previous study \citep{wang2022global}, the superposed streamlines in the present V-flame indicate no prominent recirculation region behind the annular centrebody. As a result, the V-flame dynamics more closely resemble those of amplifier flows such as a jet, resulting in increased sensitivity to acoustic perturbations \citep{schuller2003unified, schuller2003mecanismes}. Numerous studies have been dedicated to the exploration of the linear and nonlinear dynamics of V-flames under various forcing scenarios, both in confined and unconfined arrangements \citep{durox2005combustion, birbaud2007dynamics, durox2009experimental}. Self-excited oscillations have been observed in confined V-flames, which are furthermore prone to chaotic dynamics \citep{vishnu2015role}. In the case of unconfined configurations, although self-excited oscillations were observed in the experiments conducted by \cite{durox2005combustion}, a comprehensive understanding of the critical conditions and the dynamics associated with instability onset remains elusive. This present work aims to elucidate the onset of axisymmetric instability in a non-swirling annular V-flame through a combined approach of linear analysis and nonlinear time-domain simulation.    

Unsteadiness and instability in flames is a fascinating field of application for linear instability analysis because of the rich dynamics that can result from the interplay of vortical, thermal, chemical, and acoustic elements. However, reactive flows are associated with steep gradients, stiff reaction terms in the governing equations, and additional unknown variables for each chemical species, posing challenges for global linear stability analysis computations. As a result, many pioneering works invoked simplifications involving parallel flow and decoupled chemistry assumptions. For example, \citet{emerson2012} modelled the reacting wake flow behind a bluff body as an incompressible flow using discontinuous local profiles of velocity and density in the spanwise direction. Similar local analyses have also been performed on profiles measured experimentally in swirling jet flames \citep{oberleithner2015formation,douglas2021forced}, backward-facing step flames~\citep{manoharan2015absolute}, and bluff body wake flames \citep{emerson2016local}, among others. However, reacting flow fields are generally strongly non-parallel in the streamwise direction and flames interact with the flow through more than just density gradients. In recent years, there has been a gradual integration of more comprehensive global linear stability analysis methods, which preserve the streamwise variations of the base state, into canonical premixed laminar flame configurations. For example, \citet{qadri2015self} investigated the self-sustained oscillations in lifted jet diffusion flames with a heat release source term included in the governing flow equations. Using compressible reacting flow equations, \citet{avdonin2019thermoacoustic} calculated the eigenvalues and the associated eigenmode structures of a premixed slot flame, revealing significant fluctuations of streamwise velocity and chemical heat release rate along the flame surface region. In previous work, we investigated self-excited oscillations of a premixed laminar flame stabilised by a square cylinder in a channel \citep{wang2022global}. The study involved the examination of global eigenmodes associated with the steady base state of the reacting cylinder wake as well as nonlinear simulations, employing a one-step methane--air reaction in the low Mach number limit. The critical Reynolds number, corresponding to zero temporal growth rate, was identified, marking the onset of limit-cycle oscillations through a supercritical Hopf bifurcation. Endogeneity analysis \citep{Marquet2015Identifying} further indicated that this global instability was driven by momentum feedback in the wake recirculation zone, with only marginal contributions from additional feedback in the flame region, characterising the flame oscillations as a passive effect of the essentially hydrodynamic wake instability.

Other studies using global linear analysis focus on flame responses to external perturbations, including acoustic forcing, inertial waves, and optimal forcing identified through resolvent analysis. Novel reduced-order models and predictive tools for the dynamic response of flames have emerged by building upon these methods. In the case of premixed slot flames, flame transfer functions have been computed using linearised reacting flow equations with one-step \citep{avdonin2019thermoacoustic, brokof_etal_2024} and two-step chemistry schemes \citep{meindl2021spurious,wang2022linear}, and their quantitative validation against reference results has been achieved. Resolvent analysis \citep{wang2022linear} revealed that the identified optimal excitation frequency corresponds to an intrinsic thermoacoustic mode \citep{SILVA2023101065}. In the context of M-flames, computations were carried out to determine the linear response to impulsive and harmonic perturbations \citep{blanchard2015response,BlanchardPHD}. An adjoint analysis \citep{skene2019adjoint} was also employed to assess the sensitivity of the M-flames' optimal response gain to swirl. For a swirling V-flame, impulse response calculations were conducted to investigate the modulation of flame fronts by inertial waves \citep{albayrak2018response}. Recent studies have also calculated the linear responses of turbulent jet flames \citep{CASEL2022111695, kaiser2023modelling} and a reacting jet in cross flow \citep{sayadi_schmid_2021}. 

The global linear stability analysis procedure employed to study the present V-flame closely follows our previous work \citep{wang2022global} concerning a flame stabilised by a square cylinder. However, based on local stability intuition, the unsteady dynamics in these two configurations are expected to be distinct: wake recirculation as a source of local absolute instability is almost absent in the V-flame, whereas a strong convective instability resides in the jet shear layer. Moreover, the characteristic global eigenspectrum of a jet exhibits marked differences from that of a wake. In both non-reacting and reacting wakes, the global dynamics are dominated by an isolated linear eigenmode, which leads to the nonlinear shedding of counter-rotating vortices \citep{noack_eckelmann_1994,barkley2006linear,wang2022global}. In the context of round jets, such an isolated mode only exists in the presence of strong density contrast \citep{lesshafft2006nonlinear,coenen2017global,chakravarthy2018global}, where self-excited oscillations have been observed \citep{monkewitz1990self,kyle1993instability,hallberg2006universality}. In isothermal jets, the numerical eigenspectrum is typically dominated by artificial modes that arise from spurious pressure feedback between the boundaries \citep{garnaud2013modal,coenen2017global}. Such artificial modes do not converge with respect to domain size, and they are highly affected when a sponge layer is employed to artificially dampen pressure feedback \citep{cerqueira2014eigenvalue,lesshafft2018artificial}.      

Recent findings indicate that the onset of instability in a jet flows can be subcritical in many circumstances. \citet{zhu2017onset} identified a hysteretic bistable region when incrementally adjusting the jet velocity in their helium-air jet experiment. This observation suggests a subcritical Hopf bifurcation, which may be modelled through a truncated Landau equation. \citet{demange2022global} investigated the self-sustained oscillation of a heated jet with real gas effects through numerical simulation and stability analysis. The critical temperature ratio for the Hopf bifurcation, predicted by the global stability analysis around the steady base flow, was found to be considerably lower than in nonlinear simulation, indicating subcriticality. Other recent work has identified subcritical instability dynamics in constant-density, swirling circular jets~\citep{douglas_etal_2021} as well as swirling and non-swirling annular jets~\citep{douglas_etal_2022} using nonlinear branch tracing.

For the present V-flame configuration, we first conduct a global eigenmode analysis of the laminar reacting base flows shown in figure~\ref{fig:base}, obtained as steady solutions of the nonlinear reacting flow equations at various Reynolds numbers. A prominent branch of ``flame-tip'' modes, arising from non-local feedback between the nozzle and the flame tip, is discussed in particular, and non-modal behaviour is briefly characterised in the framework of resolvent analysis. Nonlinear time-stepping reveals a subcritical onset of instability, leading to periodic, quasi-periodic and chaotic regimes as the Reynolds number is varied. It is argued that, although the flow does not follow a simple supercritical Hopf bifurcation scenario, the limit-cycle is still underpinned by a similar non-local feedback mechanism as the leading linear eigenmode. Finally, a linear eigenmode analysis of the time-averaged mean flow is attempted, to assess its capability to predict the nonlinear global frequency.

The remainder of this paper is structured as follows. In \S \ref{sec:base}, we present the V-flame configuration and describe the governing equations. \S \ref{sec:linear} focuses on the global linear analysis around the steady base state, where various categories of eigenmodes are characterised. In \S \ref{sec:bifurcation}, nonlinear time stepping and analysis of the nonlinear flame dynamics are carried out. \S \ref{sec:mean} presents a linear analysis around time-averaged mean flows and discusses the ambiguity of linearised chemical reaction terms in the limit-cycle regime. Conclusions are provided in \S \ref{sec:conclusions}. 

\section{Calculation of steady V-flames}
\label{sec:base}
\noindent

\begin{figure}
    \centerline{\includegraphics[width=\textwidth]{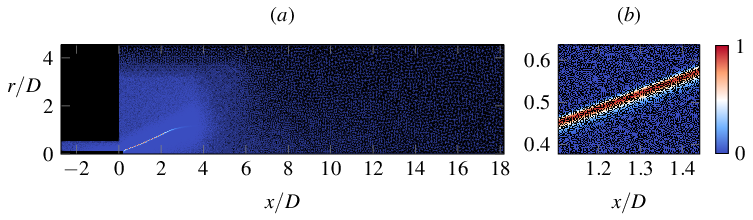}}
	\caption{The standard mesh with 180 481 elements used in the study. Colors represent the normalised base flow heat release. (\textit{a}) The complete numerical domain. (\textit{b}) Zoom on a flame surface region.}
	\label{fig:mesh}
\end{figure}

The burner geometry aligns with the numerical study by \citet{birbaud2008nonlinear} on an inverted dihedral flame, but the burner considered here is axisymmetric, formulated in the cylindrical coordinates $\boldsymbol{x}=(x, r)$. Here, $x$ represents the streamwise direction, and $r$ represents the radial direction. Figure~\ref{fig:mesh}(\textit{a}) illustrates the entire calculation domain. The streamwise position of the outer corner of the nozzle is defined as $x=0$, and the symmmetry axis is designated as $r=0$. Lean premixed methane--air reactant is injected from an annular nozzle measuring 30 mm in length and $D=11$ mm in diameter. The V-flame is anchored on a centrebody with a diameter of 3 mm, protruding 2 mm from the nozzle. The domain extends to $x_{\text{max}}=200$ mm downstream and $r_{\text{max}}=50$ mm radially. An axisymmetric condition is applied at the centerline to confine the calculation to a two-dimensional setting.

The governing equations for the reacting flow, considering an ideal gas in the low Mach number limit, are the same as in \cite{wang2022global}, except that they are formulated here in axisymmetric coordinates. Primitive variables ($\rho$, $\boldsymbol{u}$, $h$, $Y_\mathrm{CH_4}$, $p$) are chosen, where $\boldsymbol{u}=(u_x,u_r)$ represents the streamwise and radial velocity components, $\rho$ denotes density, $h$ stands for sensible enthalpy, $Y_\mathrm{CH_4}$ is the mass fraction of methane, and $p$ is the hydrodynamic pressure. The governing equations are expressed as

\begin{equation}
\label{eq:continu}
\pard{\rho}{t} = -\nabla \cdot \left (\rho \boldsymbol{u} \right ),
\end{equation}
\begin{equation}
\label{eq:momentum}
\rho\pard{\boldsymbol{u}}{t}= -  \rho \boldsymbol{u} \cdot \nabla \boldsymbol{u}   - \nabla p + \nabla \cdot \tau,
\end{equation}
\begin{equation}
\label{eq:species}
\rho\pard{ Y_\mathrm{CH_4}}{t} = -\rho \boldsymbol{u} \cdot \nabla Y_\mathrm{CH_4}+ \nabla \cdot \left (D_s \nabla Y_\mathrm{CH_4} \right ) + \dot{\omega}_\mathrm{CH_4},
\end{equation}
\begin{equation}
\label{eq:temperature}
\rho\pard{h}{t} = - \rho \boldsymbol{u} \cdot \nabla h + \nabla \cdot \left (D_h \nabla h \right ) + \dot{\omega}_T,
\end{equation}
\begin{equation}
\label{eq:perfectgas}
p_0=R_s \rho T.
\end{equation}
The molecular stress tensor $\tau$ is defined as 
\begin{equation}
\tau =\mu \left ( \nabla \boldsymbol{u} + \nabla \boldsymbol{u}^{\mathrm{T}} - \frac{2}{3} \left (\nabla \cdot \boldsymbol{u}  \right ) \boldsymbol{I} \right ), 
\end{equation}
where the molecular viscosity follows Sutherland's law $\mu=A_s T^{1/2} /(1+T_s/T)$ with constants $A_s=1.672\times10^{-6}$ $\mathrm{kg/(m.s.K^{1/2})}$ and $T_s=170.7$ K. The species and enthalpy diffusive transport coefficients, $D_s$ and $D_h$, respectively, are assumed to be proportional to the molecular viscosity, with constant values of Schmidt number $\mathrm{Sc}=\mu/D_s=0.7$ and Prandtl number $\mathrm{Pr}=\mu/D_h=0.7$. Enthalpy is expressed as $h=C_pT$, where the specific heat capacity $C_p=1.3$ $\mathrm{kJ/(kg.K)}$ is assumed constant. Following the original low-Mach number expansion of \cite{mcmurtry1986direct}, the thermodynamic pressure $p_0$ in the ideal gas law (\ref{eq:perfectgas}) is the zeroth-order pressure component, prescribed as $p_0=101.3$ kPa, while the unknown hydrodynamic pressure $p$ in the momentum equation (\ref{eq:momentum}) is its first-order complement in the squared Mach number. $R_s=264.6$ $\mathrm{J/(kg.K)}$ in the ideal gas law represents the specific gas constant. The one-step reaction scheme $\mathrm{1S\_CH4\_MP1}$ from \citet{1Step} is used, as it is sufficient to accurately reproduce the laminar burning speed of lean premixed methane--air mixtures with the current set of parameters \citep{birbaud2008nonlinear}. This scheme was also employed in our previous calculations of a bluff-body flame, successfully reproducing the length of the recirculation bubble compared to reference calculations \citep{wang2022global}. The reaction rate $\mathcal{Q}$ is governed by an Arrhenius law:
\begin{equation}
\mathcal{Q}=A_r[X_\mathrm{CH_{4}}]^{n_\mathrm{CH_4}}[X_\mathrm{O_{2}}]^{n_\mathrm{O_2}} \exp \left ( -\frac{T_{a}}{T} \right ).
\label{eq:arrhenius}
\end{equation}
For the lean global methane--air reaction modelled here, the values of the reaction exponents are $n_\mathrm{CH_4}=1$ and $n_\mathrm{O_2}=1/2$. With these exponents, the values of the Arrhenius pre-exponential factor and activation temperature are, respectively, $A_r=1.1\times10^{7}$ $\mathrm{m^{3/2}/(s.mol^{1/2})}$ and $T_a=1.007\times10^4$ K \citep{1Step}. 
The molar concentrations of $\mathrm{CH_4}$ and $\mathrm{O_2}$ are given by $[X_{\mathrm{CH_4}}]=\rho \frac{Y_{\mathrm{CH_4}}}{W_{\mathrm{CH_4}}}$ and $[X_{\mathrm{O_2}}]=\rho \frac{Y_{\mathrm{O_2}}}{W_{\mathrm{O_2}}}$, respectively, where $W_\mathrm{CH_4}=16.0$ g/mol and $W_\mathrm{O_2}=32.0$ g/mol represent their molecular masses. We assume a very lean mixture such that $Y_{\mathrm{O_2}}=0.2128$ is constant. The reaction rate in the species equation is denoted by $\dot{\omega}_\mathrm{CH_4}=-W_{\mathrm{CH_4}}\mathcal{Q}$, and the heat release due to combustion in the enthalpy equation is expressed as $\dot{\omega}_T=-\Delta h_{f}^o \mathcal{Q}$, where $\Delta h_{f}^o=-804.1$ $\mathrm{kJ/mol}$ is the standard enthalpy of reaction.

\begin{figure}
	\centerline{\includegraphics[width=0.8\textwidth]{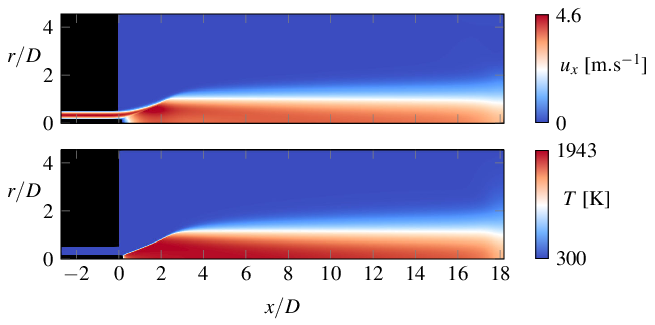}}
	\caption{Streamwise velocity and temperature associated with the steady base state at $\Rey=2282$.}
	\label{fig:base3p0}
\end{figure}

Figure~\ref{fig:base3p0} shows the base flow streamwise velocity and temperature fields at a bulk velocity of $U_0=3.0$ $\mathrm{m.s^{-1}}$. For this condition, the corresponding inflow Reynolds number defined as $\Rey = \rho_0 U_0 D/\mu_0$ is equal to 2282. A simple parabolic profile is prescribed as the inlet velocity. The inflow conditions for temperature and fuel mass fraction are set to $T_0=300$ K and $Y_{\mathrm{CH_4}}=0.04256$. No-slip and adiabatic conditions are imposed on the inflow channel walls. The dump-plane wall ($x=0$) connected to the outer corner of the annular nozzle is assumed to be no-slip and isothermal at $T_0=300$ K. Note that the use of a slow co-flow, as employed for stabilisation purposes in \citet{birbaud2008nonlinear}, is not considered here. No-slip and isothermal conditions are also imposed on the annular centrebody, with the wall temperature fixed at $T_r=700$ K, as in the reference. At the lateral boundary far from the flame ($r=r_{\mathrm{max}}$), a free-slip condition is imposed on the velocity, and the temperature and fuel mass fraction are set to $T_0=300$ K and $Y_{\mathrm{CH_4}}=0$, respectively. A symmetry condition is imposed along the centreline. Finally, a traction-free condition is employed at the downstream boundary ($x=x_{\mathrm{max}}$), with Neumann conditions on the temperature and fuel mass fraction. 

In the following, we investigate V-flames in a range of bulk velocities from $U_0=2.2$ $\mathrm{m.s^{-1}}$ to $U_0=3.8$ $\mathrm{m.s^{-1}}$. The prescribed conditions result in $\Rey=\rho_0 U_0 D/\mu_0$ ranging from 1674 to 2891, where $\rho_0$ and $\mu_0$ denote the density and molecular viscosity at the inflow. Figure~\ref{fig:base} shows the base flame shape and vorticity distribution at both ends of the Reynolds number range investigated. The flame is elongated downstream as the velocity is increased. 

\begin{table} 
	\caption{Sponge layer parameters and number of mesh elements for eigenmode calculations in figure~\ref{fig:eigs_sponge_convergence}.}
	\begin{center}
		\begin{tabular}{c c c c c c} 
			\hline
			Case &Marker & $x_{\text{sg}}/D$ & $r_{\text{sg}}/D$ & $\alpha^{-1}$ & Mesh elements\\ 
			\hline
			($a$) &+ & 13.6 & 2.7 & 20 & 180 481\\ 
			\hline	
			($b$) &$\triangle$& 10.9
			& 2.3 & 40 & 180 481\\ 
			\hline	
			($c$) &$\circ$& 9.1
			& 1.8 & 60 & 180 481\\
			\hline	
			($d$) &$\diamond$& 13.6 & 2.7 & 20 & 221 019\\
			\hline	
			($e$) &$\square$& 13.6 & 2.7 & 20 & 251 757\\
		\end{tabular}
	\end{center}
	
	\label{tab:legend}
\end{table}

Sponge layers are introduced at the downstream ($x=x_{\mathrm{max}}$) and lateral boundaries ($r=r_{\mathrm{max}}$) to prevent spurious back-scattering~\citep{lesshafft2018artificial}. The sponge layer implementation from \citep{meliga2010effect} is employed, wherein viscosity is artificially increased in the sponge regions. The molecular viscosity $\mu$ is divided by the sponge layer expression $sg(r,x)$, formulated as
\begin{equation}
\begin{aligned}
sg(r,x) &= 1 \text{  if } r\le r_{\text{sg}} \text{ and } x\le x_{\text{sg}},\\
sg(r,x) &= 1+(\alpha-1) \zeta(x,x_{\text{sg}}) \text{  if } r\le r_{\text{sg}} \text{ and } x > x_{\text{sg}},\\
sg(r,x) &= sg(r_{\text{sg}},x)+\left[\alpha-sg(r_{\text{sg}})\right]\zeta(r,r_{\text{sg}})  \text{  if } r > r_{\text{sg}},
\end{aligned}
\end{equation}
where $\alpha^{-1}$ is listed in table~\ref{tab:legend}. Here, $x_{\text{sg}}$ and $r_{\text{sg}}$ denote the starting position of the sponge layers in the streamwise and radial directions. The function $\zeta$, defined by
\begin{equation}
\zeta(a,b)=\frac{1}{2}+\frac{1}{2}\tanh \left\{\gamma \tan \left(-\frac{\pi}{2}+\pi \frac{|a-b|}{l}  \right)  \right\},
\end{equation} 
yields a smooth transition of $\mu$ from the inner region to the truncated boundaries. In this expression, the constant $\gamma$ is set to 1 and $l$ denotes the sponge layer thickness, equal to $x_{\text{max}}-x_{\text{sg}}$ or $r_{\text{max}}-r_{\text{sg}}$ in the streamwise and radial directions, respectively. It is important to note that the transport coefficients $D_s$ and $D_h$ are also increased in the sponge layer as they are proportional to $\mu$. This implementation is applied both in the computation of the base flow and in the subsequent linear analysis.

The nonlinear governing equations are discretised on the unstructured mesh presented in figure~\ref{fig:mesh} using 180 481 triangular elements. High grid resolution is allocated at the flame surface, as shown in figure~\ref{fig:mesh}(\textit{b}), with a maximum mesh resolution of $\Delta x=0.08$ mm. A continuous Galerkin method is employed using mixed Taylor--Hood finite elements, of quadratic order for the velocity and linear order for other flow variables. The base states are calculated by Newton iteration. Readers may refer to Appendix B of \citet{wang2022phd} for more details on the numerical methods.

\section{Global linear analysis}

\label{sec:linear}

\subsection{Survey of the global eigenspectrum}

To analyse the onset and behaviour of self-sustained oscillations in V-flames, a global linear stability analysis is conducted by computing the eigenmodes associated with the Jacobian matrix of the governing equations, which are linearised around the steady base state. Equations (\ref{eq:continu}-\ref{eq:temperature}) are linearised with respect to the primitive variables ($\rho$, $\boldsymbol{u}$, $h$, $Y_\mathrm{CH_4}$, $p$) at each grid point of the entire numerical domain. The flow fluctuations $\boldsymbol{q'}(\boldsymbol{x},t)$ can be expanded in the basis of eigenmodes $\boldsymbol{q}'_j(\boldsymbol{x},t)=\boldsymbol{\phi}_j(\boldsymbol{x})\exp(\lambda_j t)$ obtained from the generalised eigenvalue problem,
\begin{equation}
\label{eq:direct}
\lambda_j \boldsymbol{B}\boldsymbol{\phi}_j =\boldsymbol{A}\boldsymbol{\phi}_j.
\end{equation}
The matrices $\boldsymbol{B}$ and $\boldsymbol{A}$ are constructed from the linearisation of (\ref{eq:continu}-\ref{eq:temperature}). The eigenvalues $\lambda_j$ and associated eigenvectors $\boldsymbol{\phi}_j$ are computed using the \texttt{eigs()} function in Matlab. The frequency is defined as $2\pi f_j=\text{Im}[\lambda_j]$, and the growth rate is defined as $2\pi \sigma_j=\text{Re}[\lambda_j]$, so that $\lambda_j/(2\pi)=\sigma_j+i f_j$. Furthermore, their non-dimensional counterparts are defined as the Strouhal number $\mathrm{St}=\frac{fD}{U_0}$ and $\mathrm{\tilde{\sigma}}=\frac{\sigma D}{U_0}$. Instabilities associated with non-zero azimuthal numbers are not considered in this study, although they are known to arise spontaneously in isothermal annular jets~\citep{douglas_etal_2022} or in conical flames~\citep{douglas_etal_2023}.

\begin{figure} 
	\centerline{\includegraphics[width=\textwidth]{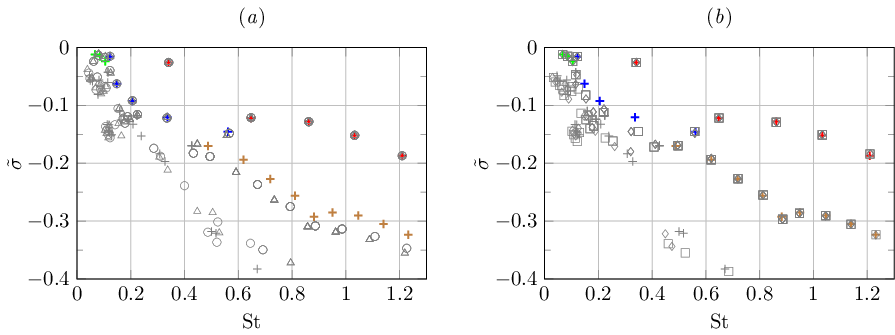}}
	\caption{Eigenspectra at $\Rey=2282$. (\textit{a}) Convergence with respect to parameters of sponge layers. (\textit{b}) Convergence with respect to grid resolution. Legends are given in table \ref{tab:legend}. Different categories of eigenmodes are marked with colors: flame-column modes (green ``${\color{green}+}$'') in figure~\ref{fig:eigs_others}(\textit{a}-\textit{c}), plume modes (blue ``${\color{blue}+}$'') in figure~\ref{fig:eigs_others}(\textit{d}-\textit{f}), flame-tips modes (red ``${\color{red}+}$'') in figure~\ref{fig:eigs_tip} and flame-tip-downstream modes (brown ``${\color{brown}+}$'') in figure~\ref{fig:eigs_others}(\textit{g}-\textit{i}).}
	\label{fig:eigs_sponge_convergence}
\end{figure}

The linear eigenmodes associated with steady base states at Reynolds numbers ranging from $\Rey=1674$ to $\Rey=2891$ with an increment of $\Delta \Rey=76$ are calculated (corresponding to an inflow velocity increment of $\Delta U_0=0.1$ $\mathrm{m.s^{-1}}$). Different families of eigenmodes are identified based on their dependence on boundary conditions, their eigenmode structures, and their trends with respect to the Reynolds number. Specifically, their convergence with respect to parameters of the sponge layers is examined, by using the different values given in rows ($a$-$c$) of table~\ref{tab:legend}. From case ($a$) to ($c$), the sponge layers are increasingly large and viscous. The sponge layer in case ($c$) corresponds to around one-half of the whole domain length, with a sixty-fold increased molecular viscosity at the boundaries. The results are presented in figure~\ref{fig:eigs_sponge_convergence}($a$), where the eigenmodes that overlap with the three different markers are considered converged.

\begin{figure} 
	\centerline{\includegraphics[width=\textwidth]{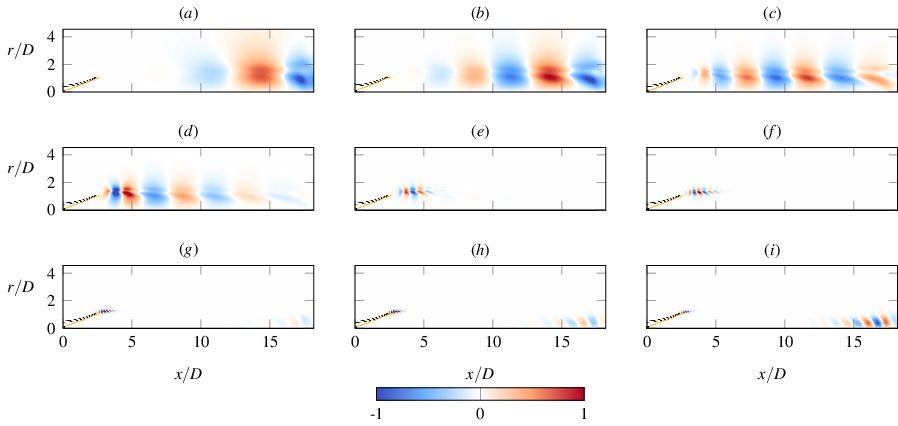}}
	\caption{Three families of eigenmodes at $\Rey=2282$. Radial velocity fluctuations are shown. (\textit{a}-\textit{c}) Flame column modes: first three modes with green markers ``{\color{green}+}'' in figure~\ref{fig:eigs_sponge_convergence}. (\textit{d}-\textit{f}) Plume modes: first three modes with blue markers ``{\color{blue}+}''. (\textit{g}-\textit{i}) flame-tip-column modes: first three modes with brown markers ``{\color{brown}+}''. Black contour: an isocontour of base state vorticity illustrating the free shear layer position; yellow contour: an isocontour of base state heat release rate illustrating the flame front position. }
	\label{fig:eigs_others}
\end{figure}

\begin{figure} 
	\centerline{\includegraphics[width=0.9\textwidth]{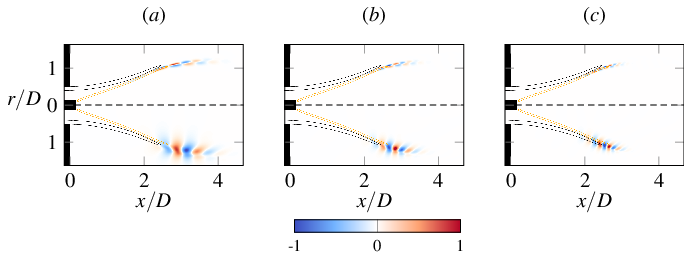}}
	\caption{Branch of flame tip modes at $\Rey=2282$. First three modes (\textit{a}-\textit{c}) with red markers ``{\color{red}+}'' in figure~\ref{fig:eigs_sponge_convergence}. Top of each plot: heat release rate fluctuation. Bottom of each plot: radial velocity fluctuation. Black contour: an isocontour of base state vorticity illustrating the free shear layer position; yellow contour: an isocontour of base state heat release rate illustrating the flame front position. }
	\label{fig:eigs_tip}
\end{figure}

\begin{figure} 
	\centerline{\includegraphics[width=\textwidth]{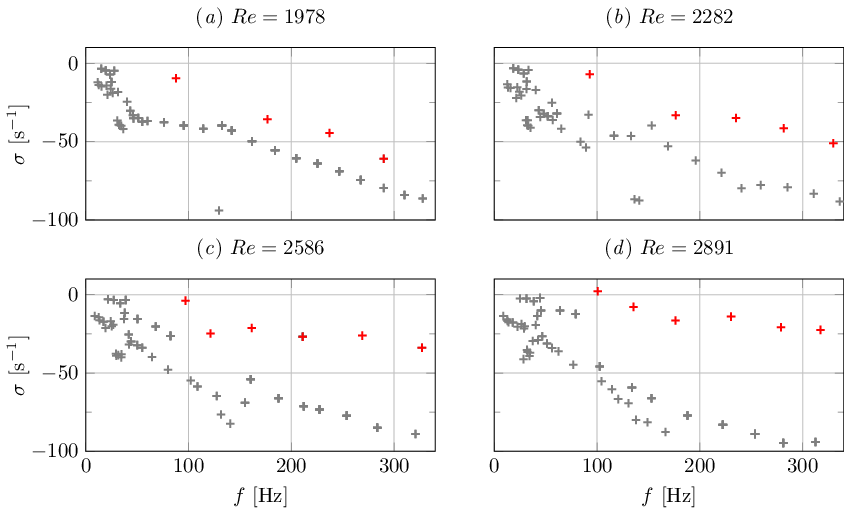}}
	\caption{Eigenspectra at different Reynolds number. Branches of flame-tip modes are marked as red. The dimensional temporal growth rates and frequencies are used for comparison.}
	\label{fig:eigs_Re}
\end{figure}

The eigenfunctions associated with the identified categories are shown in figures \ref{fig:eigs_others} and \ref{fig:eigs_tip}, marked with different cross colors in figure~\ref{fig:eigs_sponge_convergence}($a$), using the parameters of case ($a$) in table~\ref{tab:legend}. In each figure, the position of the flame surface is illustrated by a yellow isocontour corresponding to 10\% of the maximum heat release rate, while the shear is illustrated by a black isocontour corresponding to 33\% of the maximum azimuthal vorticity. The eigenmode spectra at Reynolds numbers equal to 1978, 2282, 2586, and 2891 are presented in figure~\ref{fig:eigs_Re}. Four main families of eigenmodes are identified:

\begin{itemize}
	\item[(i)] \emph{Flame-column modes} (green crosses ``${\color{green}+}$'' in figure~\ref{fig:eigs_sponge_convergence}, eigenmodes shown in \ref{fig:eigs_others}\textit{a}-\textit{c}). These modes are characterised by their low frequencies; they are the least stable family of eigenmodes for $\Rey<2586$. These modes do not undergo destabilisation as the Reynolds number increases but are notably influenced by the parameters of the sponge layer. The eigenmode structures exhibit spatial growth primarily in the streamwise direction. Specifically, the mode associated with the lowest frequency, illustrated in figure~\ref{fig:eigs_others}(\textit{a}), displays maximum amplitude at the downstream end of the computational domain. This fluctuation expands radially to both sides of the shear layer, with the inner side extending close to the centerline. At higher frequencies, the peak fluctuation moves upstream, and spreads closer to both sides of the shear layer. Such modes, characterised by an amplitude maximum near the domain's downstream end, are commonly encountered in incompressible and compressible jets, where they are attributed to the stable advection of nearly neutral structures within the shear layer \citep{garnaud2013modal}. In general, these modes do not converge with larger domain sizes and are influenced by the sponge layers at the boundary \citep{lesshafft2018artificial}. Given that the fluctuations fill the column of the plume, we denote these modes as flame-column modes, akin to jet-column modes.         
	
	\item[(ii)] \emph{Plume modes} (blue crosses ``${\color{blue}+}$'' in figure~\ref{fig:eigs_sponge_convergence}, eigenmodes associated with the three lowest frequencies shown in \ref{fig:eigs_others}\textit{d}-\textit{f}). Along this branch, identified at relatively low frequencies, the temporal growth rates exhibit an increasing trend with frequency. Remarkably, these modes remain unaffected by the sponge layer, suggesting that they form a family distinct from the flame column modes. Although the branch experiences a slight destabilisation with growing Reynolds numbers, all modes within this range of $\Rey$ remain stable. The fluctuation structures are primarily localised slightly downstream of the flame surface and within the jet shear region, corresponding to areas of heightened base vorticity. Notably, the maximum perturbation is consistently observed around $x/D=4$ for the three leading eigenmodes, positioned closely along the shear in the radial direction. This specific location corresponds to the point where the base flow velocity and temperature profiles develop to be parallel in the streamwise direction, creating a base flow reminiscent of a non-reacting hot jet. Similar eigenmode structures were previously identified in the mixing layer of a plume, as illustrated in figure~4(\textit{b}) of \citet{chakravarthy2018global}. In their study of plumes, the maximum fluctuation amplitude is found very close to the inflow, corresponding to the region with the maximum density gradient in the base flow and confined within the mixing layer. Despite the absence of buoyancy effects in the present governing equations, we refer to this branch as plume modes. 

	\item[(iii)] \emph{Flame-tip modes} (red crosses ``${\color{red}+}$'' in figure~\ref{fig:eigs_sponge_convergence}, leading eigenmodes shown in figure~\ref{fig:eigs_tip}, red crosses ``${\color{red}+}$'' in figure~\ref{fig:eigs_Re}). This branch constitutes a prominent feature in V-flames, notably separated from other more stable eigenvalues in the spectra. The eigenmodes along this branch remain unaffected by the sponge layer. However, with an increase in Reynolds number, certain members of this branch become unstable. Specifically, the mode associated with the lowest frequency becomes unstable at $\Rey=2815$, suggesting a Hopf bifurcation. Subsequently, we refer to this mode as the ``leading flame-tip mode.'' The fluctuation of heat release rate and velocity is confined to the inner mixing layer, extending closely downstream from the intersection of the flame surface and the jet shear. The maximum fluctuation is identified in the flame extinction zone at around $x/D=3$ for the leading flame-tip mode. For higher-frequency flame-tip modes, the maximum is located further upstream, closer to the flame surface, and the associated fluctuations exhibit higher wavenumbers.

	\item[(iv)] \emph{Flame-tip-column modes} (brown crosses ``${\color{brown}+}$'' in figure~\ref{fig:eigs_sponge_convergence}, eigenmodes shown in \ref{fig:eigs_others}\textit{g}-\textit{i}). This branch encompasses stable eigenmodes across a broad range of frequencies. The eigenmodes exhibit fluctuations at the flame extinction region and downstream close to the centreline. Notably, the fluctuation amplitudes at the downstream end become more pronounced with higher frequency. However, these modes are sensitive to the sponge layer parameters and remain unaffected by changes in Reynolds number.
\end{itemize}

The convergence of eigenmodes with respect to the mesh refinement is examined at $\Rey=2282$ using meshes with 180 481, 221 019 and 251 757 elements. Figure~\ref{fig:eigs_sponge_convergence}(\textit{b}) shows that convergence has been achieved on the reference mesh with 180 481 elements for all the eigenmodes described above except certain plume modes. The poor convergence of the plume modes seems to result from the fact that the maximum fluctuation of those plume modes are located at around $x/D=4$ where the change of local refinement is relatively abrupt. Nonetheless, as the least stable plume mode is converged with respect to refinement on the reference mesh and its associated growth rate is negative, we conclude that further refinement is not necessary for our global instability study.

To summarise, the flame-column modes and the flame-tip-column modes exhibit intense oscillations at the end of the numerical domain, with their associated eigenvalues significantly influenced by sponge layer parameters. This suggests that these modes arise from spurious pressure feedback from the outflow, rather than from physical instabilities. In contrast, the plume and the flame-tip modes remain unaffected by the sponge layers, indicating a more physical nature. The plume modes appear to result from the extended shear layer located downstream from $x/D=4$, sharing similar base flow and eigenmode structures with a non-reacting plume. Conversely, the flame-tip modes, characterised by strong oscillations at the flame extinctions around $x/D=3$, become considerably more destabilised as the Reynolds number increases, leading to a positive growth rate at $\Rey=2891$. The leading flame-tip mode, \textit{i.e.}, the least stable mode along the flame-tip branch, demonstrates an increased frequency with the Reynolds number, as shown in figure~\ref{fig:eigs_Re}.

\subsection{Analysis of the flame-tip mode mechanisms}\label{sec:mechanisms}

We now aim to characterise the physical mechanisms governing the behaviour of the leading flame-tip modes, drawing inspiration from earlier experimental work by \citet{schuller2003mecanismes} and \citet{durox2005combustion}. In their studies of perturbed laminar V-flames, the flame oscillations were interpreted as a consequence of vortex structures, advected from the burner exit along the shear layer, modulating the flame shape. Building upon this interpretation, the time delay between the heat release rate and an imposed velocity perturbation at the inlet can be linked to the convective velocity of the vortices and the convection distance. Experimental measurements of this time delay were performed, and the convective velocity was estimated as one-half of the maximum streamwise velocity at the burner exit. The convection distance was then determined by multiplying the time delay by the convective velocity, revealing a correspondence with a region where the roll-up of the flame surface was notably pronounced.

In contrast to these acoustically-forced experimental investigations, our focus is specifically toward the frequency of the leading flame-tip eigenmode, which is representative of intrinsic flame dynamics in the absence of external forcing. When the leading eigenmode is unstable, self-excited flame oscillations are expected. In our approach, we posit that the eigenfrequency of a flame-tip mode in a V-flame is linked to the advection time from the nozzle exit to the flame surface along the outer shear layer. The assumption of linearity underlines that the base flow remains unaffected by perturbations, ensuring steady streamlines and a steady flame shape in the base state. This allows for a more precise calculation of the convective velocity and distance compared to the finite-amplitude (i.e. nonlinear) oscillating flame scenarios explored experimentally by \citet{schuller2003mecanismes} and \citet{durox2005combustion}. However, the calculation is not without challenges, primarily due to the following factors. Firstly, the start- and endpoints of the convective distance are not clearly defined. Secondly, the shear layer exhibits a continuum of advection velocities across its finite thickness, and the precise streamline along which the vortices are advected remains to be chosen. Finally, hydrodynamic perturbations are generally dispersive, and are not advected by the base flow in a completely passive manner.

\begin{figure} 
	\centerline{\includegraphics[width=0.9\textwidth]{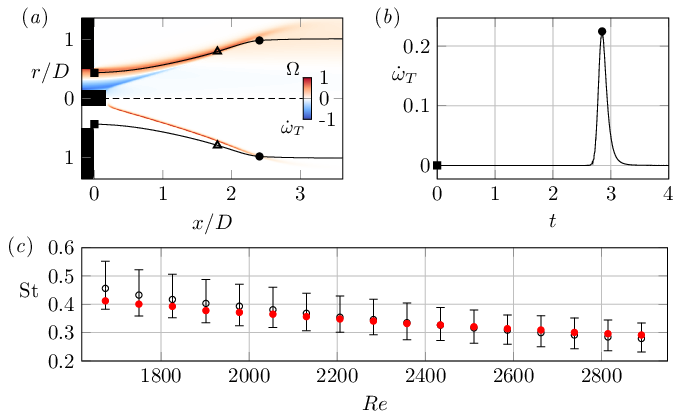}}
	\caption{Calculation of Lagrangian advection time and comparison with leading flame-tip mode frequencies. (\textit{a}) Steady base states at $\Rey=2282$. Top: vorticity $\Omega$. Bottom: heat release rate $\dot{\omega}_T$. Both variables are normalised with their maximum. Marker ``$\triangle$'' represents the point associated with maximum vorticity. Marker ``$\blacksquare$'' represents the intersection of backward integration and $x=0$. Marker ``$\bullet$'' represents the peak of Lagrangian heat release rate along the streamline stemming from ``$\blacksquare$''. (\textit{b}) Lagrangian heat release as a function of time $t$ normalised by $D/U_0$. The advection time is obtained as the time lag between ``$\blacksquare$'' and ``$\bullet$''. (\textit{c}) Evolution of the Strouhal number with Reynolds number. Marker ``${\color{red}\bullet}$'' denotes the leading flame-tip modes. Marker $\circ$ represents the Strouhal number calculated from the average advection time over all starting grid points where the vorticity exceeds 99\% of its global maximum magnitude. The associated error bars are shown.}
	\label{fig:convection_time}
\end{figure}

To address these challenges, we rely on two key assumptions in our calculation. First, we presume that the eigenmode oscillation stems from a non-local interaction between known up- and downstream points. The upstream interaction initiates at the nozzle exit ($x=0$), while the downstream interaction concludes at the point where the streamline intersects with the flame front. Second, we assume that the perturbations in the shear layer traverse through the region marked by the highest base flow vorticity without dispersion. Based on these assumptions, the advection time is computed from the base flow in three sequential steps:

\begin{itemize}
	\item[(i)] We identify the points corresponding to the strongest vorticity in the shear layer, denoted by ``$\triangle$'' in figure~\ref{fig:convection_time}(\textit{a}). 
	\item[(ii)] A backward integration is executed from ``$\triangle$'' based on the base flow velocity field. The starting point where the obtained streamline intersects with the burner exit at $x=0$ is then determined and marked as ``$\blacksquare$''.
	\item[(iii)] A forward integration is initiated from ``$\blacksquare$'', through ``$\triangle$''. Along this trajectory, we record the Lagrangian heat release rate, as illustrated in figure~\ref{fig:convection_time}(\textit{b}). The intersection between the streamline and the flame surface is defined by the peak of the Lagrangian heat release rate. This point is interpreted as the downstream terminus of the non-local interaction and is marked as ``$\bullet$''. The advection time is subsequently determined as the time lag along the streamline between ``$\blacksquare$'' and ``$\bullet$''.
\end{itemize}

This process is repeated for all points in the shear layer where the vorticity exceeds 99\% of its global maximum magnitude. Consequently, a collection of differing advection times is obtained for each base flow at a specified Reynolds number. In figure~\ref{fig:convection_time}(\textit{c}), the black dots depict the averaged values of the Strouhal number, calculated as the inverse of the advection time, with error bars indicating the maximum and minimum within each set. The red dots represent the Strouhal number associated with the leading flame tip eigenmode. Remarkably, over the extensive range of Reynolds numbers investigated, the frequency of the leading flame tip mode closely aligns with the Lagrangian advection time. It is crucial to again note that the calculation does not consider the dispersion of perturbations, and instead assumes that perturbations travel at the base flow velocity. This assumption, along with spatial variations in the effective centre of the up- and downstream interaction regions, may explain the small errors observed at certain Reynolds number ranges.

This outcome suggests that the identified flame-tip modes indeed originate from a non-local feedback between the nozzle exit and the flame surface with a frequency characterised by the advection time of downstream traveling hydrodynamic perturbations along the outer shear layer. It can be hypothesised that the non-local feedback loop is closed by upstream-travelling pressure waves generated by the fluctuations at the flame tip. In the low Mach number limit, this pressure impulse is felt instantaneously at the nozzle outlet.

\begin{figure} 
	\centerline{\includegraphics[width=0.9\textwidth]{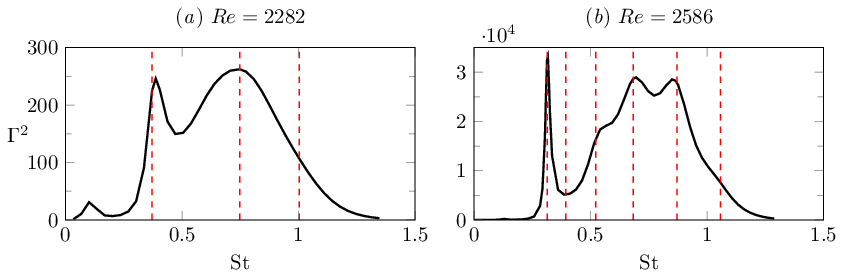}}
	\caption{Optimal gain curves at $\Rey=2282$ (\textit{a}) and $\Rey=2586$ (\textit{b}) obtained by the resolvent analysis. The red dashed lines represent the frequency of the flame-tip eigenmodes in figure~\ref{fig:eigs_Re}.}
	\label{fig:xy_resolvent}
\end{figure}

\begin{figure} 
	\centerline{\includegraphics[width=0.6\textwidth]{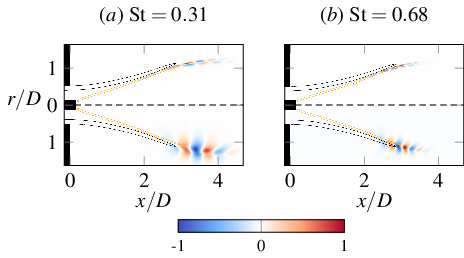}}
	\caption{Resolvent response structures at $\Rey=2586$ for $\mathrm{St}=0.31$ (\textit{a}) and $\mathrm{St}=0.68$ (\textit{b}). Top of each plot: heat release rate fluctuation. Bottom of each plot: radial velocity fluctuation. Black contour: an isocontour of base state vorticity illustrating the free shear layer position; yellow contour: an isocontour of base state heat release rate illustrating the flame front position.}
	\label{fig:para_resolvent}
\end{figure}

\subsection{Resolvent analysis}

When the flame is subjected to external perturbations, its linear response may be dominated by pseudo-resonance \citep{trefethen2005spectra}, owing to the non-normality of the operator. This possibility is assessed through resolvent analysis, performed on the base flow at $\Rey=2282$ and $\Rey=2586$, with volume forcing in the axial and the radial momentum equations. The response is calculated using the same boundary conditions as in the global eigenspectrum calculations, and the standard 2-norm of velocity is employed both for the forcing and response norms \citep{garnaud2013preferred}. The optimal gain $\Gamma^2$ relating the forcing and response norms at different Strouhal numbers $\mathrm{St}$ are presented in figure~\ref{fig:xy_resolvent}, where the frequencies corresponding to the flame-tip eigenmodes in figure~\ref{fig:eigs_Re}(\textit{b},\textit{c}) are illustrated as the red dashed lines. At both Reynolds numbers, the dominant resolvent gain peaks align well with flame-tip mode frequencies. Figure~\ref{fig:para_resolvent} presents the optimal response structures at the local peak frequencies $\mathrm{St}=0.31$ and $\mathrm{St}=0.68$ related to $\Rey=2586$, which are characterised by structures that are clearly similar to the flame-tip eigenmodes displayed in figure~\ref{fig:eigs_tip}. These results indicate that the linear flame dynamics are dominated by modal mechanisms, even in the presence of flow forcing.

\section{Nonlinear analysis}
\label{sec:bifurcation}

\begin{figure} 
	\centerline{\includegraphics[width=0.9\textwidth]{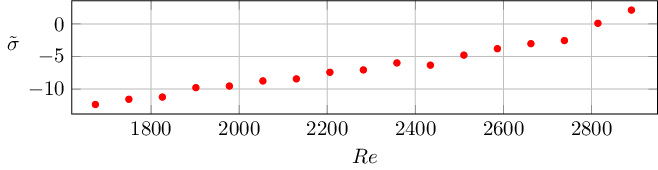}}
	\caption{Normalised temporal growth rate of the leading flame-tip mode with Reynolds number.}
	\label{fig:sigma_Re}
\end{figure}

\begin{figure} 
	\centerline{\includegraphics[width=\textwidth]{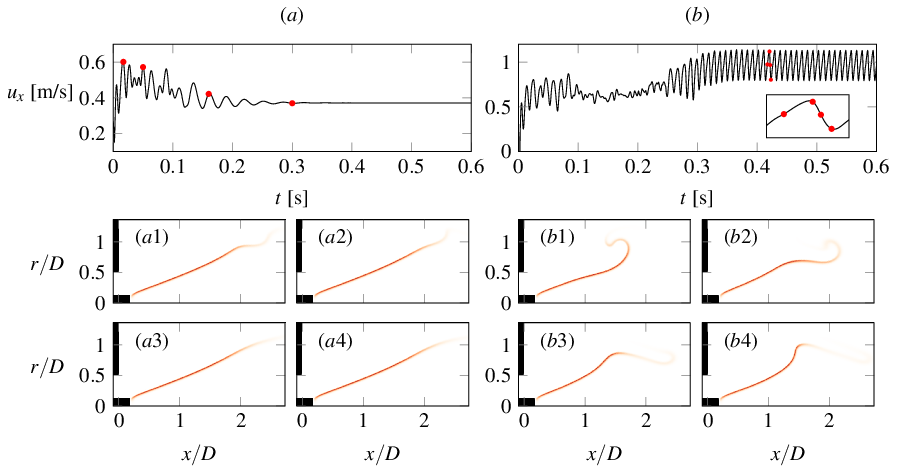}}
	\caption{Illustration of subcritical behaviour at $\Rey=1978$. Streamwise velocity signal recorded at $(x/D,y/D) = (2.7,1.2)$ with initial velocity perturbations of (\textit{a}) $\beta=20\%$ and (\textit{b}) $\beta=50\%$ magnitude. The snapshots of flame movement are shown in (\textit{a}1-\textit{a}4) and (\textit{b}1-\textit{b}4), corresponding to the time steps marked with red dots ``{\color{red}$\boldsymbol{\cdot}$}''. A zoom-in of the time series is given in (\textit{b}) over one period of the limit-cycle oscillation.}
	\label{fig:animation}
\end{figure}

\subsection{Time-series analysis}
\noindent 
The outcomes derived from the linear analysis indicate a Hopf bifurcation at $\Rey=2815$, as evidenced by the temporal growth rate evolution associated with the flame-tip modes in figure~\ref{fig:sigma_Re}. In this section, we delve into the nonlinear dynamics of the V-flame by superimposing finite-amplitude perturbations to the steady base states at each $\Rey$ and conducting nonlinear time integration using a first-order implicit scheme. The results depicted in figure~\ref{fig:animation} correspond to a linearly stable base state at $\Rey=1978$. We denote the unperturbed base flow velocity $\boldsymbol{u}_b (\boldsymbol{x})$. The superposed perturbations are prescribed as the velocity components of the leading flame-tip eigenmode, represented by $\boldsymbol{u}_p(\boldsymbol{x})$. Both components of the velocity perturbation are normalised such that their maximum is the same maximum amplitude associated with the streamwise base flow velocity. The perturbed initial velocity field $\boldsymbol{u}_0 (\boldsymbol{x})$ can be expressed as:
\begin{equation}
	\boldsymbol{u}_0 (\boldsymbol{x})=\boldsymbol{u}_b (\boldsymbol{x})+\beta\boldsymbol{u}_p(\boldsymbol{x}),
\end{equation}
where $\beta$ is a ratio constant to be varied. We consider two perturbation amplitude ratios of $\beta=20\%$ and $\beta=50\%$. In each scenario, the temporal signal of the streamwise velocity is recorded at $(x/D,r/D) = (2.7,1.2)$ in a region proximal to the flame extinction zone. Figure~\ref{fig:animation} also presents snapshots of the flame surface at the time steps marked with red dots in the time signal.

At the initial perturbation amplitude of $\beta=20\%$, the velocity perturbation undergoes transient growth before it enters modal decay. Only a small section of the flame surface near the downstream edge ($x/D>2$) displays visible oscillations during the transient growth, after which the flame surface converges to the steady base state. This transient growth is an indicator of the strong non-normality of the system, which may promote bypass to other sub- or non-critical attractors, as in classical Poiseuille and Couette flows \citep{schmid2002stability}. Indeed, at a larger initial amplitude of $\beta=50\%$, the perturbation undergoes growth before settling into a limit-cycle oscillation. Four snapshots in an oscillation cycle are presented in figure~\ref{fig:animation}(\textit{b}1-\textit{b}4). A considerable portion of the flame surface ($x/D>1$) exhibits oscillations, resulting in pronounced wrinkling along the length of the flame as well as dramatic flapping of the flame tip, resembling the oscillating flame surfaces observed in \citet{durox2005combustion}. The results indicate that the system has at least two distinct nonlinear attractors at $\Rey=1978$.

\begin{figure} 
	\centerline{\includegraphics[width=0.9\textwidth]{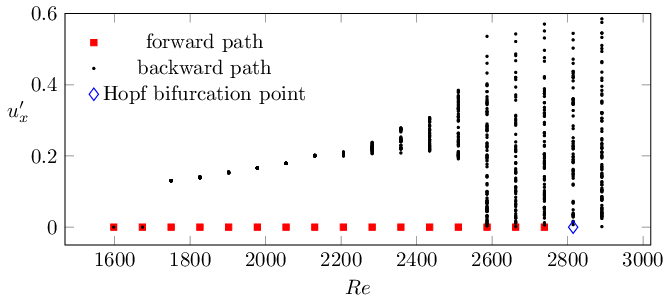}}
	\caption{Bifurcation diagram. The backward continuation path is represented with black dots ($\boldsymbol{\cdot}$), corresponding to the local maxima of streamwise velocity perturbation signal measured. The stable range of forward continuation path is represented by red squares (${\color{red}\blacksquare}$). The Hopf bifurcation point of the steady state is represented by a blue diamond (${\color{blue}\diamond}$).   }
	\label{fig:dyn_bif}
\end{figure}

\begin{figure} 
	\centerline{\includegraphics[width=\textwidth]{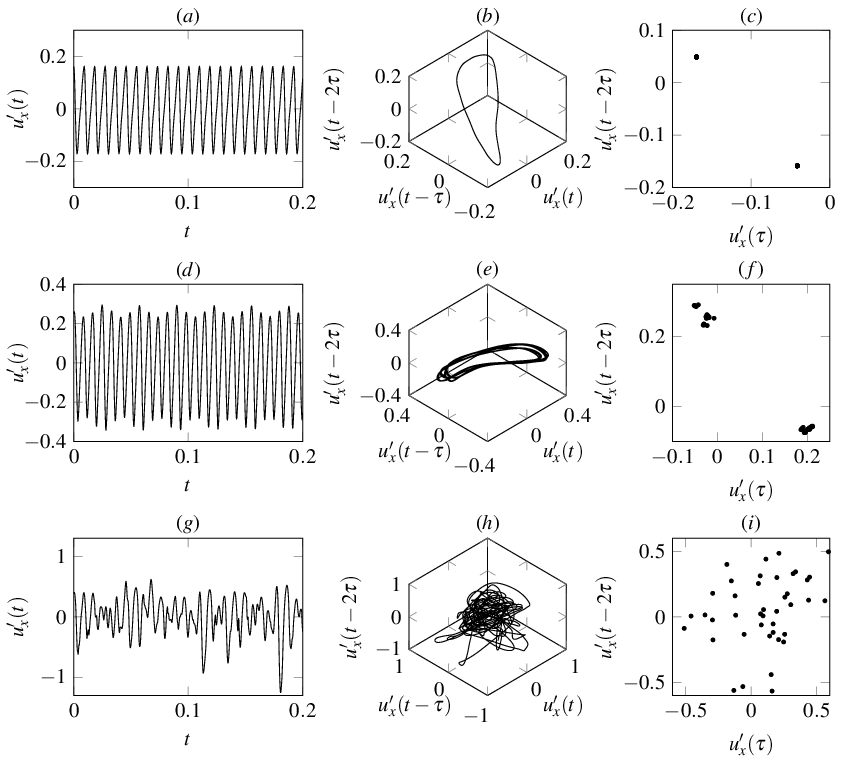}}
	\caption{(\textit{a}, \textit{d}, \textit{g}) Time signal of streamwise velocity perturbation measured at $(x/D,r/D) = (2.7,1.2)$. (\textit{b}, \textit{e}, \textit{k}) 3D Poincaré sphere. (\textit{c}, \textit{f}, \textit{i}) Poincaré section. (\textit{a}, \textit{b}, \textit{c}) $\Rey=1978$. (\textit{d}, \textit{e}, \textit{f}) $\Rey=2358$. (\textit{g}, \textit{h}, \textit{i}) $\Rey=2815$.   }
	\label{fig:Poincare}
\end{figure}

The bifurcation diagram with respect to the Reynolds number is depicted in figure~\ref{fig:dyn_bif}, revealing the presence of a bi-stable region between $\Rey=1750$ and $\Rey=2815$. Both forward and backward continuation paths, characterised by the streamwise velocity perturbation $u'_x$, are displayed in the figure, and the methods for tracking both paths are distinct. 

\noindent
\emph{Forward path}: the forward continuation path along the steady base state is traced by introducing small-amplitude velocity fluctuations of the leading flame-tip mode onto the steady states at each investigated Reynolds number. The initial perturbation is set to the velocity of the leading eigenmode with a small amplitude of 1\% of the maximum base flow streamwise velocity. For $\Rey<2815$, all flames disturbed in this way reconverge to the original steady states, akin to the time series in figure~\ref{fig:animation}(\textit{a}). These steady states are represented by the red squares (${\color{red}\blacksquare}$) in figure~\ref{fig:dyn_bif}. At $\Rey=2815$, marked by the blue diamond (${\color{blue}\diamond}$), the disturbed flame undergoes temporal growth before entering an unsteady fluctuating state. A similar dynamic process is observed for the disturbed steady flame at $\Rey=2891$. At both Reynolds numbers, the simulation duration corresponds to around six flow-through times of the computational domain including the sponge layer. This long duration was necessary to ensure subsidence of the initial transient, such that the fluctuations converge to a statistically stationary state. The resulting loss of stability of the steady state apparent from the time-domain simulations aligns with the critical $\Rey$ for instability identified via global linear analysis in \S~\ref{sec:linear}.

\noindent
\emph{Backward path}: the backward continuation path is identified by gradually reducing the Reynolds number with $\Delta \Rey=76$, starting with the unsteady fluctuating state at the next-highest $\Rey$. (For example, the $\Rey=2815$ case is initialised with the final state from the $\Rey=2891$ case). The black dots ($\boldsymbol{\cdot}$) in figure~\ref{fig:dyn_bif} represent any local maxima identified in the velocity fluctuation signal at the probe location $(x/D,r/D) = (2.7,1.2)$. The cut-off time horizon before recording the local maxima corresponding to three flow-through times for the Reynolds number $2130 \leq  \Rey \leq 2815$. The values of the local maxima are normalised by the temporal average of the velocity time series, representing the relative fluctuation amplitudes. The results show that fluctuations persist even when $\Rey$ is decreased below the critical Reynolds number to $\Rey=2663$. This indicates that the system exhibits subcritical dynamics and possesses at least two attractors below the critical point. The distribution of local maxima markedly narrows when the Reynolds number is reduced to 2510, indicating a change in the dynamic state. The distribution then further narrows when reducing the Reynolds number from 2510 to 2130. At $\Rey=2130$, the distribution of local maxima collapses essentially to a single value, with the local maxima varying by less than 1\% of the sliding average of ten successive local maxima samples. The presence of a single local maxima value is consistent with the limit-cycle oscillation behaviour observed in figure~\ref{fig:Poincare}(\textit{a}). Once the Reynolds number is reduced below 1750, the self-sustained oscillations vanish, and the system converges again to the steady state, as shown by the collapse of red squares and black dots at $\Rey=1598$ and $\Rey=1674$, the lower limit of our investigation.

The dynamic states associated with the oscillating flame at $\Rey=1978$, $\Rey=2358$ and $\Rey=2815$ are further characterised by plotting the temporal signal of $u'_x(t)$ at $(x/D,r/D) = (2.7,1.2)$, the associated 3-D Poincaré trajectory computed with a time delay of $\tau=5\times 10^{-3}$ s and the corresponding 2-D Poincaré section in figure~\ref{fig:Poincare}. Figure~\ref{fig:Poincare}(\textit{b-c}) illustrate that the associated phase space at $\Rey=1978$ is an enclosed trajectory with two points depicted in the intersection plane, indicating a limit-cycle state. At $\Rey=2358$, the temporal signal displays a nearly enclosed trajectory and plane intersection points scattered in clusters, a pattern representative of quasi-periodic dynamics. Finally, unlike the organised state-space nature identified for the lower Reynolds numbers, the temporal signal for $\Rey=2815$ reveals erratic and intermittent behaviours, corresponding to the local maxima in figure~\ref{fig:dyn_bif} being densely distributed from zero to the maximum amplitude. In particular, the Poincaré section of figure~\ref{fig:Poincare}(\textit{f}) exhibits scattered points that suggest a chaotic nature of the flame fluctuations. Overall, this progression is consistent with a Reulle--Takens--Newhouse scenario for the onset of chaos in the annular V-flame.

\begin{figure} 
	\centerline{\includegraphics[width=0.75\textwidth]{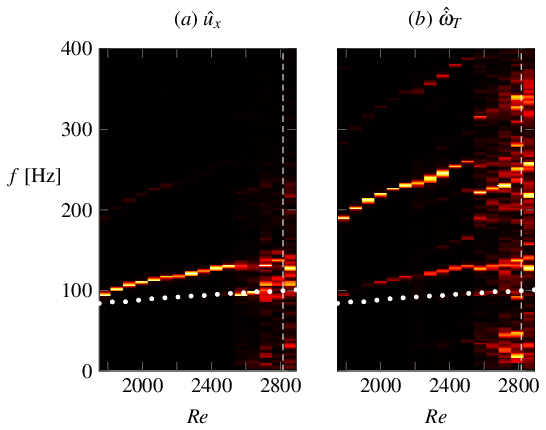}}
	\caption{Power density spectra along the backward continuation path by tracking (\textit{a}) local streamwise velocity and (\textit{b}) the global heat release rate. The frequencies of the leading linear flame-tip mode are superposed as white dots. The white dashed lines represent the Reynolds number associated with Hopf bifurcation. The spectra are displayed in the Reynolds number range $1750 \leq  \Rey \leq 2891$. No flame oscillation is observed for $\Rey \leq 1674$.}
	\label{fig:PSD}
\end{figure}

Power Spectral Density (PSD) contours, computed along the backward continuation path, are depicted in figure~\ref{fig:PSD}, tracking the axial velocity perturbation at $(x/D, r/D) = (2.7,1.2)$ and the global heat release rate. The abscissa ranges from $\Rey=1750$, the lower extent of the identified bi-stable region, to the upper limit represented by the critical $\Rey$ at the white dashed line. The linear flame-tip mode frequencies obtained from the imaginary part of the leading linear eigenvalue is overlaid on the PSD. Tracking the point velocity evolution reveals a single fundamental frequency peak for $\Rey<2586$, distinct from the flame-tip modes. At $\Rey=2586$, a second incommensurate frequency peak, closely aligned with the linear flame-tip mode frequency, emerges. This suggests that a 2-torus is born in the phase space via a Neimark--Sacker bifurcation, and the system has transitioned from periodicity now to two-frequency quasiperiodicity. In contrast to the two-frequency quasiperiodicity presented in a previous forced synchronisation study \citep{guan2019open}, where quasiperiodicity arises from the competition between a self-excited mode and a forced mode, this one arises from the competition between two incommensurate self-excited modes. Further increases in Reynolds number result in more continuously distributed frequency spectra, indicative of the non-periodic state observed at $\Rey>2586$. Examining the global heat release rate unveils dominant frequency components at twice the frequency observed when tracking point velocity. This indicates nonlinear harmonic interactions -- a notable feature of the flame dynamics that cannot be captured by linear analysis. 

In summary, increasing $\Rey$ along the forward continuation path confirms that even small-amplitude perturbations lead to unsteady oscillations above the critical Reynolds number identified through linear analysis. Subcritical dynamics corresponding to hysteresis is identified along the backward continuation path when gradually reducing the Reynolds number, resulting in progressive transitions among unsteady dynamic states before the end of the hysteresis interval. At $\Rey=2510$, close to the critical Reynolds number at $\Rey=2815$, the dominant nonlinear oscillation frequency peak aligns with the linear flame-tip mode frequency, though other frequencies remain present. The nonlinear frequency associated with the subcritical oscillation below $\Rey=2510$ is apparently unrelated to the flame-tip mode frequencies identified in the linearly stable subcritical base flows. A qualitative interpretation of the subcritical dynamics can be hypothesised from the flame shape snapshots in figure~\ref{fig:animation}: nonlinear oscillations are self-sustaining when a sufficiently large portion of the downstream flame edge is perturbed with sufficiently large amplitude and vice versa.

\subsection{Analysis of the nonlinear non-local interaction}
\label{sec:nonlinear_nonlocal}

Having already explored the linear dynamics in \S~\ref{sec:mechanisms}, we here investigate the physical mechanics of the subcritical limit-cycle oscillation at $\Rey=1978$ in the nonlinear case. As in figure~\ref{fig:PSD}, the system exhibits a strong limit cycle oscillation in this regime with a fundamental frequency of $f_0=110.0$ Hz at $\Rey=1978$. Informed by the linear non-local mechanism explored above, we hypothesise a nonlinear non-local scenario where vortex structures advected from the burner exit interact with the flame surface. Thus, as before, we posit that the limit-cycle frequency can be estimated by the convection time from the burner exit to the flame surface. However, unlike the linear case, in the nonlinear scenario, the unsteady flame surface exhibits substantial movement. Hence, the convective time delay between when a vortex structure departs from the burner exit and when it arrives at the flame surface depends strongly on the starting phase, denoted as $\phi_{\mathrm{start}}=2\upi{}t_{\mathrm{start}}/T_0$, where $T_0=1/f_0$ is the oscillation period and $t_{\mathrm{start}}$ is the starting time instant. The value of convective time delay from the burner exit at $x=0$ to the downstream flame surface also depends on the exact starting radial position of a measured trajectory, denoted as $r_{\mathrm{start}}$. 

We conduct a Monte Carlo simulation to generate pathlines initiated at different $t_{\mathrm{start}}$ and $r_{\mathrm{start}}$. A numerical non-inertial particle tracer solver is implemented and coupled with the nonlinear flame solver. The tracer trajectories are computed using the same value of discretisation time as the flame solver, and the particle is advanced using a fourth-order Runge-Kutta scheme. The starting time $t_{\mathrm{start}}$ is considered at each time step over one cycle period $T_0$, resulting in 92 points of phase $\phi_{\mathrm{start}}$. Regarding $r_{\mathrm{start}}$, we prescribe a distribution centred at $r=0.436D$, corresponding to the mean radial position of trajectories experiencing the strongest vorticity magnitudes identified in the linear regime. The prescribed $r_{\mathrm{start}}$ distribution covers the range from $0.409D$ to $0.464D$ with 31 points. Hence, the overall number of sampling points for the Monte Carlo simulation is $N_t\times N_r = 92\times31=2852$.

The obtained $N_t\times N_r$ passive tracer trajectories are analysed. The convective time delay from the burner exit to the flame surface, denoted as $\Delta t_{\mathrm{conv}}$, is identified as the difference between the peak instant of Lagrangian heat release rate and the starting time in the same manner as in \S~\ref{sec:mechanisms}. To identify the most relevant tracer trajectories, a selection process is conducted, based on an assumption that only the trajectories experiencing sufficient magnitudes of vorticity and heat release rate along the trajectories are essential to the global flame dynamics. The vorticity magnitude criterion is designed to select trajectories corresponding to the advection of vortex structures, while the heat release rate criterion serves to select trajectories that actually cross the flame surface. A vorticity magnitude threshold $\Omega_{\mathrm{th}}$ and a heat release rate threshold $\dot{\omega}_{T,\mathrm{th}}$ are defined, respectively, to select the trajectories and estimate the oscillation frequency $f_e$. A formula for the estimation of $f_e$ is proposed as:
\begin{equation}
	\label{eq:estimate}
	f_e=\frac{N_t N_r}{\sum_{i=1}^{N_t} \sum_{j=1}^{N_r} \delta_{i,j}},
\end{equation}   
where
\begin{equation}
	\delta_{i,j}=\begin{cases}
		& \Delta t_{\mathrm{conv},i,j}, \text{ if } \max(\dot{\omega}_{T,i,j})>\dot{\omega}_{T,\mathrm{th}} \text{ and } \max(\Omega_{i,j})>\Omega_{\mathrm{th}}, \\ 
		& 0, \text{ else.}
	   \end{cases}
\end{equation}
The formula implies that $f_e$ is simply estimated as the inverse of the mean convective time delay of the selected trajectories. The indices $i,j$ represent the variables of a trajectory of the sampling space with $i \in [1,N_t]$ and $j \in [1,N_r]$. $\max(\dot{\omega}_{T,i,j})$ refers to the maximum heat release rate along the entire trajectory, whereas $\max(\Omega_{i,j})$ refers to the maximum vorticity magnitude along the trajectory portion $x/D>0.3$. (The latter restriction prevents large vorticity magnitudes localised at the burner lip from interfering with the identification of advecting vortex structures.)

\begin{figure} 
	\centerline{\includegraphics[width=0.8\textwidth]{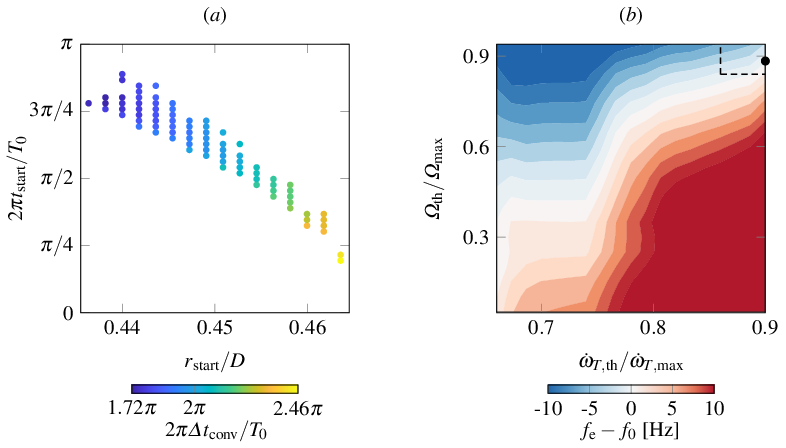}}
	\caption{(\textit{a}) Selected trajectories using a set of threshold $\Omega_{\mathrm{th}}=0.89 \Omega_{\max}$ and $\dot{\omega}_{T,\mathrm{th}}=0.94 \dot{\omega}_{T,\max}$. Each trajectory is associated with a set of starting instant and starting radial position ($t_\mathrm{start}$, $r_\mathrm{start}$), coloured by the value of convective time delay $\Delta t_\mathrm{conv}$. (\textit{b}) Differences between the estimated frequency $f_e$ and the actual nonlinear frequency $f_0$ using different sets of selection thresholds ($\dot{\omega}_{T,\mathrm{th}}$, $\Omega_{\mathrm{th}}$). Differences $|f_e-f_0|$ larger than 10 Hz are represented by the same degree of red or blue colours. The black circle marker corresponds to the selection criterion used in (\textit{a}). The dashed rectangle in the top right corner represents the trajectories filtered by the highest values of $\dot{\omega}_{T,\mathrm{th}}$ and $\Omega_{\mathrm{th}}$ where $|f_e-f_0|<5$ Hz.}
	\label{fig:para_MC}
\end{figure}

Figure~\ref{fig:para_MC}(\textit{a}) illustrates the selected sampling points corresponding to a prescribed set of threshold $\Omega_{\mathrm{th}}=0.89 \Omega_{\max}$ and $\dot{\omega}_{T,\mathrm{th}}=0.94 \dot{\omega}_{T,\max}$, where $\Omega_{\max}$ and $\dot{\omega}_{T,\max}$ represent the maximum vorticity magnitude and heat release rate recorded among all the trajectories of the selected sampling points. Each point is associated with a set of ($t_\mathrm{start}$, $r_\mathrm{start}$), coloured by the value of convective time delay $\Delta t_\mathrm{conv}$. The trajectories at smaller $r_\mathrm{start}$ (i.e. close to the annular jet core) experience smaller convective time delays before reaching the flame surface, and vice versa. 

The estimated frequency $f_e$ obtained by Eq.~(\ref{eq:estimate}) on different sets of threshold values ($\dot{\omega}_{T,\mathrm{th}}$, $\Omega_{\mathrm{th}}$) is presented in figure~\ref{fig:para_MC}(\textit{b}). The contour colours indicate the difference between $f_e$ and the reference frequency $f_0=110.0$ Hz extracted from the nonlinear simulation. The bottom right of this figure corresponds to high values of $\dot{\omega}_{T,\mathrm{th}}$ and low values of $\Omega_{\mathrm{th}}$. Hence, this region selects trajectories closer to the jet core, where tracers are more likely to reach the flame surface but are also likely to experience smaller vorticity. As noted above, due to the higher fluid velocities near the jet core, these trajectories also experience smaller $\Delta t_\mathrm{conv}$ and therefore larger values of $f_e$. Conversely, the top left part of figure~\ref{fig:para_MC}(\textit{b}) corresponds to low values of $\dot{\omega}_{T,\mathrm{th}}$ and high values of $\Omega_{\mathrm{th}}$. This region tends to contain trajectories starting near the burner wall with higher vorticity but also with less likelihood to intersect the flame surface at a point where the heat release is large. Likewise, since these trajectories are initialised in the slow-moving fluid near the wall, they feature longer convective time delays and lower values of $f_e$. The trajectories with high values of both $\dot{\omega}_{T,\mathrm{th}}$ and $\Omega_{\mathrm{th}}$ lead to a relatively small difference between $f_e$ and $f_0$. More specifically, the absolute difference $|f_e-f_0|$ obtained over the range $\dot{\omega}_{T,\mathrm{th}}/\dot{\omega}_{T,\mathrm{max}} \in [0.79,0.94]$ and $\Omega_{\mathrm{th}}/\Omega_{\mathrm{max}} \in [0.86,0.90]$ is less than 5 Hz within the dashed rectangle on the top right corner of figure~\ref{fig:para_MC}(\textit{b}). (Further increase in the direction of either $\dot{\omega}_{T,\mathrm{th}}$ or $\Omega_{\mathrm{th}}$ leads to a sharp decline in the number of available sampling points, leading to deviated values of $f_e$.) Hence, the trajectories that yield the best frequency estimates according to \eqref{eq:estimate} both (1) exhibit prominent vortex structures and (2) intersect the flame surface where and when the local instantaneous heat release rate is large. This supports our hypothesis that these processes are essential to the flame dynamics.

\begin{figure} 
	\centerline{\includegraphics[width=0.55\textwidth]{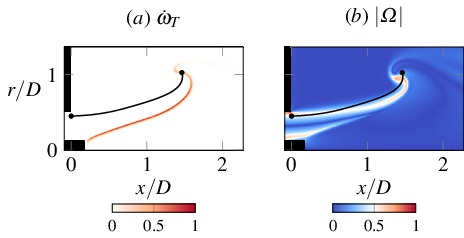}}
	\caption{Snapshots of heat release rate (\textit{a}) and vorticity magnitude (\textit{b}) at the time instant of mean intersection time $t_\mathrm{int}$ corresponding to the selected trajectories in figure~\ref{fig:para_MC}(\textit{a}). The trajectory at $2\pi{}t_{\mathrm{start}}/T_0=0.30$ and $r_{\mathrm{start}}/D=0.45$ is superposed.}
	\label{fig:para_endtime}
\end{figure}

The sampling points illustrated in figure~\ref{fig:para_MC}(\textit{a}) correspond to the black circle marker in figure~\ref{fig:para_MC}(\textit{b}) at high threshold values of $\dot{\omega}_{T,\mathrm{th}}$ and $\Omega_{\mathrm{th}}$. Using these points as input for \eqref{eq:estimate}, $f_e$ is identified as 109.8 Hz, very close to the reference value of $f_0=110.0$ Hz. The time instant when a trajectory intersects the flame surface can be calculated as $t_\mathrm{int}=t_\mathrm{start}+\Delta{}t_\mathrm{conv}$. We find that the standard deviation of $t_\mathrm{int}$ calculated from the different trajectories in figure~\ref{fig:para_MC}(\textit{a}) is only $2.4\%$ of a cycle period, indicating that these trajectories reach the flame surface at nearly the same time instant. Figure~\ref{fig:para_endtime} presents the snapshot of heat release rate and vorticity magnitude at the time instant associated with the mean value of $t_\mathrm{int}$. Comparing this snapshot against the snapshots in figure~\ref{fig:animation}, we note that the point where the pathlines associated with the strongest vorticity intersect the flame surface is situated at a position relatively close to the upstream limit of the flame flapping motion. Further, a structure of significant vorticity magnitude can be visualised close to to the flame tip in figure~\ref{fig:para_endtime}(\textit{b}). The trajectory at $t_{\mathrm{start}}/T_0=0.30$ and $r_{\mathrm{start}}/D=0.45$, corresponding to $\Delta{}t_{\mathrm{conv}}/T_0=0.99$ is superposed in figure~\ref{fig:para_endtime}. Note that this pathline trajectory does not correspond to any streamline of the instantaneous flow field, because the flow is unsteady. Compared with the base flow streamline identified within linear regime in figure~\ref{fig:convection_time}(\textit{a}), the axial position where the trajectory intersects the flame surface is significantly reduced. The result aligns with the higher oscillation frequency $f_0=110.0$ Hz observed in the nonlinear regime in comparison to the linear flame-tip mode identified at 87.8 Hz.

\section{Global linear analysis of the nonlinear mean flow}
\label{sec:mean}
\noindent

\begin{figure} 
	\centerline{\includegraphics[width=0.4\textwidth]{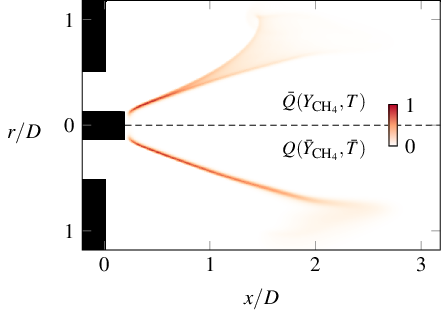}}
	\caption{Time-averaged reaction rate $\mathcal{Q}(Y_{\mathrm{CH_4}},T)$ (top) and reaction rate of time-averaged variables $\mathcal{Q}(\bar{Y}_{\mathrm{CH_4}},\bar{T})$ (bottom). The variables are normalised by their maximum.  }
	\label{fig:Q_mean}
\end{figure}

Finally, we compute the global eigenmodes of the time-averaged mean flow to investigate the potential recovery of the nonlinear oscillation frequency and/or structure by these mean flow eigenmodes. From the outset, this procedure is complicated by the non-unique definition of a mean flow, exposed by \citet{karban2020ambiguity}, that arises from nonlinearity. Those authors demonstrated the issue by comparing resolvent analysis results for a compressible jet, obtained from averaging the same LES data in either primitive or conservative variables. The present reacting flow case provides an even more compelling illustration of mean-flow ambiguity. We will consider two equally plausible definitions of the mean reaction rate $\mathcal{Q}$, noting that many more are possible. Representing the time average by an overbar, the first definition is the average of $\mathcal{Q}$ itself,   
\begin{equation}
	\bar{\mathcal{Q}}(X_i, T)=\overline{A_r[X_\mathrm{CH_{4}}]^{n_\mathrm{CH_4}}[X_\mathrm{O_{2}}]^{n_\mathrm{O_2}} \exp \left ( -\frac{T_{a}}{T} \right )}.
	\label{eq:arrhenius_time}
\end{equation}
The second definition inserts the mean flow variables into the definition of $\mathcal{Q}$,  
\begin{equation}
	\mathcal{Q}(\bar{X}_i, \bar{T})=A_r[\bar{X}_\mathrm{CH_{4}}]^{n_\mathrm{CH_4}}[\bar{X}_\mathrm{O_{2}}]^{n_\mathrm{O_2}} \exp \left ( -\frac{T_{a}}{\bar{T}} \right ).
	\label{eq:arrhenius_model}
\end{equation}
The mean reaction rate, assessed by both definitions, is depicted in figure~\ref{fig:Q_mean}. $\bar{\mathcal{Q}}(X_i, T)$ reveals pronounced oscillations of the flame surfaces, leading to the progress rate being distributed in a region around flame extinction. Conversely, $\mathcal{Q}(\bar{X}_i, \bar{T})$ displays only a thin flame surface and fails to represent the unsteady oscillations. It is worthwhile to note that in turbulent reacting flows, the difference between two reaction rates is often employed to assess the turbulence--chemistry interaction, which is important in turbulent reaction modelling \citep{poinsot2005theoretical,duan2011assessment,di2021direct}. Conventionally, the second definition $\mathcal{Q}(\bar{X}_i, \bar{T})$, called the \textit{laminar chemistry} or \textit{laminar reaction rate} model, only takes the frozen mean flow quantity into account, as if in a steady laminar flame. Conversely, the first definition $\bar{\mathcal{Q}}(X_i, T)$, called the \textit{turbulent reaction rate}, includes the species and temperature fluctuations modulated by the turbulent flow field. Figure~\ref{fig:Q_mean} shows that even in a laminar flame, unsteadiness can lead to significant differences between the mean reaction rates evaluated by these definitions. Thus, the conventional notion of a \textit{laminar chemistry} model requires caution.

In the linearised reacting flow equations, the expression for the linearised reaction rate, $\mathcal{Q}'$, is derived as
\begin{equation}
	\mathcal{Q}'=\bar{\mathcal{Q}}\left (  (n_\mathrm{CH_4}+n_\mathrm{O_2})\frac{\rho'}{\bar{\rho}}+n_\mathrm{CH_4}\frac{Y'_\mathrm{CH_{4}}}{\bar{Y}_\mathrm{CH_{4}}}+ n_\mathrm{O_2}\frac{Y'_\mathrm{O_{2}}}{\bar{Y}_\mathrm{O_{2}}}  +\frac{T_a T'}{\bar{T}^2}   \right ),
	\label{eq:arrhenius_linear}
\end{equation}
where the mean reaction rate $\bar{\mathcal{Q}}$ is factored out and provided by the second expression $\mathcal{Q}(\bar{X}_i, \bar{T})$. Note that, in accordance with our fuel-limited reaction model, $Y'_{\mathrm{O_2}}=0$ by assumption.

\begin{figure} 
	\centerline{\includegraphics[width=0.8\textwidth]{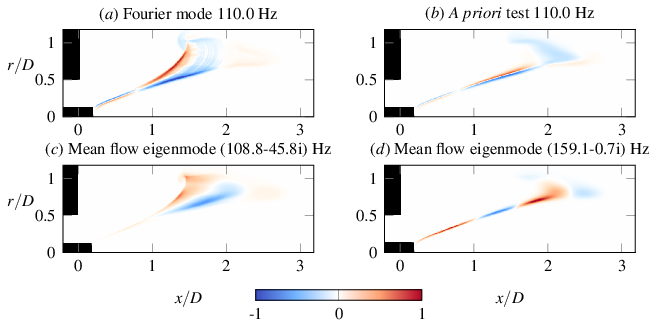}}
	\caption{Reaction rate fluctuation $\mathcal{Q}'$ at $\Rey=1978$. (\textit{a}) Fourier modes at the limit-cycle fundamental frequency of 110.0 Hz, obtained from nonlinear simulation. (\textit{b}) \textit{A-priori} test at 110.0 Hz. (\textit{c}) Computed mean flow eigenmode corresponding to the eigenvalue at $(108.8 -45.8\mathrm{i})$ Hz. (\textit{d}) Computed mean flow eigenmode corresponding to the eigenvalue at $(159.1 -0.7\mathrm{i})$ Hz.}
	\label{fig:Q_fluc}
\end{figure}

\begin{figure} 
	\centerline{\includegraphics[width=0.5\textwidth]{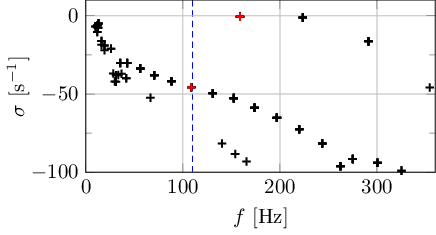}}
	\caption{Eigenspectrum of the time-averaged mean flow at $\Rey=1978$. Blue dashed line indicates the limit-cycle frequency of the velocity perturbation at 110.0 Hz. Eigenvalues marked in red are $(108.8 -45.8\mathrm{i})$ Hz and $(159.1 -0.7\mathrm{i})$ Hz, for which the corresponding eigenmode structures are given in figure~\ref{fig:Q_fluc} and \ref{fig:u_fluc}, respectively.}
	\label{fig:mean_eigs}
\end{figure}

\begin{figure} 
	\centerline{\includegraphics[width=0.5\textwidth]{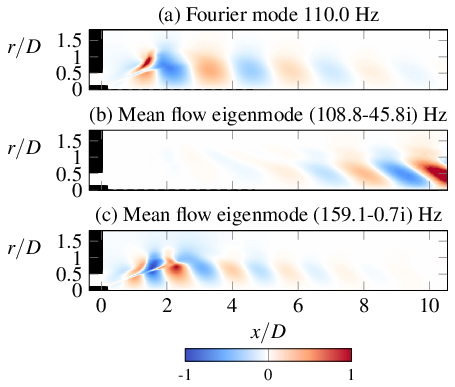}}
	\caption{Radial velocity fluctuation $u'_r$ at $\Rey=1978$. (\textit{a}) Fourier modes at the limit-cycle fundamental frequency of 110.0 Hz, obtained from nonlinear simulation. (\textit{b}) Computed mean flow eigenmode corresponding to the eigenvalue at $(108.8 -45.8\mathrm{i})$ Hz. (\textit{c}) Computed mean flow eigenmode corresponding to the eigenvalue at $(159.1 -0.7\mathrm{i})$ Hz.}
	\label{fig:u_fluc}
\end{figure}

A mean flow analysis is conducted at $\Rey=1978$, where the saturated unsteady dynamics correspond to a subcritical limit-cycle oscillation. The Fourier mode is extracted at the limit-cycle fundamental frequency of 110.0 Hz, as identified in figure~\ref{fig:Poincare}(\textit{a-c}) and~\ref{fig:PSD}. Figure~\ref{fig:Q_fluc}(\textit{a}) presents the associated Fourier mode of progress rate fluctuation, serving as a reference for fluctuation structures. These structures manifest as progress rate wrinkles convecting along the distributed mean reaction zone, as assessed through $\bar{\mathcal{Q}}(Y_{\mathrm{CH_4}},T)$. Concurrently, figure \ref{fig:u_fluc}(\textit{a}) illustrates the Fourier mode of radial velocity, revealing oscillations in both the flame region and the extended shear layer. The identified Fourier mode structures underscore the distinctive characteristics of subcritical nonlinear dynamics in the V-flame compared to the linear flame-tip modes. Note that the extracted frequency 110.0 Hz is not the dominant frequency of the heat release rate, which appears at the first harmonic (\textit{cf}. Figure~\ref{fig:PSD}(b)).

The outcomes of the \textit{a priori} assessments for $\mathcal{Q}'$ utilising $\bar{\mathcal{Q}}=\mathcal{Q}(\bar{Y}_{\mathrm{CH_4}},\bar{T})$ in its formulation are displayed. Fourier modes of the flow variables $Y'_\mathrm{CH{4}}$ and $T'$ are inserted into (\ref{eq:arrhenius_linear}), following the methodology employed for \textit{a priori} tests in turbulent reaction models \citep{kaiser2023modelling}. The structure of $\mathcal{Q}'$ acquired using $\bar{\mathcal{Q}}=\mathcal{Q}_(\bar{Y}_{\mathrm{CH_4}},\bar{T})$ is depicted in figure~\ref{fig:Q_fluc}(\textit{b}), showing travelling waves along the thin flame surface. These structures are akin to the shape of mean flow chemical progress rates; however, they deviate markedly from the reference Fourier mode structures.

The resulting mean flow spectrum presented in figure~\ref{fig:mean_eigs} reveals a separated branch of eigenvalues, akin to the flame-tip modes identified for the base flow in \S~\ref{sec:base}. The two leading eigenvalues on this branch exhibit temporal growth rates close to zero, but their associated frequencies are distinct from the 110.0 Hz fundamental frequency of the nonlinear limit-cycle state as indicated by the dashed blue line. The mean flow eigenmode structures associated with the eigenvalues marked with red are also displayed. Of these, one occurs at $(108.8 -45.8\mathrm{i})$ Hz, a heavily damped eigenvalue with a frequency close to the fundamental tone of the nonlinear oscillation. Another is the leading, marginally-damped eigenvalue on the separated branch at $(159.1 -0.7\mathrm{i})$ Hz. The mean flow modes of progress rate fluctuation in figure~\ref{fig:Q_fluc}(\textit{c-d}) and those of radial velocity fluctuation in figure~\ref{fig:u_fluc}(\textit{b-c}) were found not to align with the reference Fourier modes. The strong velocity fluctuations downstream in figure~\ref{fig:u_fluc}(\textit{b}) indicates that the eigenvalue $(108.8 -45.8\mathrm{i})$ Hz is similar to a flame-column mode, and it is not relevant to the fundamental frequency of the nonlinear oscillation. The structure of velocity fluctuation of the leading eigenvalue $(159.1 -0.7\mathrm{i})$ Hz in figure~\ref{fig:u_fluc}(\textit{c}) shares a certain resemblance to the Fourier mode in figure~\ref{fig:u_fluc}(\textit{a}), although the identified eigenvalue frequency is notably different.

At this point, we emphasise that linear eigenmode analysis of the time-averaged mean flow is not expected in principle to predict the nonlinear limit-cycle frequency.  For example, \citet{sipp2007global} have shown through weakly-nonlinear analysis that its success depends on whether nonlinear interactions are strongly dominated by mean flow distortion effects, provided these effects are stabilising (i.e. that the bifurcation is supercritical). Building on this work and others~\citep[e.g.][]{karban2020ambiguity,beneddine2016conditions,tammisola2016coherent}, the failure of mean flow analysis in the present V-flame can be attributed to its particular instability dynamics which have been identified and analysed in the previous sections.

Foremost, when the relevant bifurcation is subcritical, there should be no expectation of validity for mean flow analysis even if the \citet{sipp2007global} criterion on the Stuart--Landau coefficients is satisfied. Unlike the supercritical case, subcritical dynamics indicate that harmonic interactions induce a fundamentally destabilising effect in the vicinity of the bifurcation (i.e. for sufficiently small amplitudes). This may be because mean flow distortion is destabilising or unsteady harmonic feedback is destabilising, or both. In subcritical systems, saturation does not become possible until the amplitude reaches some finite values, indicating that harmonic interactions play a different role on the limit cycle in comparison to the base state. The ``real-zero imaginary-frequency'' (RZIF) property of a mean flow \citep{turton2015prediction} results from the growth of an unstable linear eigenmode that gradually distorts the mean flow until it can grow no further; clearly, in a subcritical situation where linear growth is impossible, that scenario cannot play out.

Additionally, the oscillations identified in the limit-cycle regime are strongly dichromatic, and the non-local instability mechanism crucially depends on interactions between features associated with both frequencies. While there is no reason why a global mean flow analysis could not deal with non-local interactions on their own, the unsteady heat release rate in this case cannot be accurately represented at the fundamental frequency, as evidenced by the \textit{a priori} test in figure~\ref{fig:Q_fluc}(\textit{b}). Indeed, as is common of premixed flames exhibiting strong flame--flow interactions~\citep[see, for example, the seminal observations of][]{joos1986selfexcited,lang1991harmonic}, the global heat release rate oscillates predominantly at the first harmonic of the fundamental velocity oscillation tone (see figure~\ref{fig:PSD}). This harmonic generation phenomenon can be understood intuitively, since the global heat release is proportional to the flame surface area, which increases twice per cycle -- at both extremes of a flow oscillation~\cite{lieuwen2005nonlinear}. Conversely, as discussed by \citet{tammisola2016coherent}, linear analysis assumes that all fields (including the heat release rate) oscillate at a single global frequency. Hence, the harmonic interaction between the heat release rate and flow velocity in oscillating premixed flames can not be captured by linear mean flow analysis -- this interaction is fundamentally nonlinear.

A strong intrinsic nonlinearity of the mean reaction rate was also encountered in our prior work on flames anchored in the wake of a 2-D square cylinder in a channel~\citep{wang2022global}. However, the mean flow analysis in that work accurately captured the oscillation frequency. This difference can be mainly attributed to the distinct instability mechanisms in a cylinder wake flame and an annular V-flame. The dominant dynamics in the former case arise from a supercritical hydrodynamic shear instability (the Bénard--von Kármán instability) localised in the wake recirculation zone, with only secondary influences from the spatially-separated flame front region. Conversely, the investigated V-flame dynamics is subcritical and there is a significant non-local flame--flow interaction involving two distinct frequencies, as discussed in \S~\ref{sec:nonlinear_nonlocal}.

\section{Conclusions}
\label{sec:conclusions}

This study computationally investigates the self-excited axisymmetric oscillations of a lean premixed V-flame in a laminar annular jet. The reactive flow is simulated using an irreversible single-step global chemistry model representing a lean premixed methane--air reaction coupled to the low-Mach number compressible Navier--Stokes equations. Following the identification of steady states of the linearised reacting flow equations, we conduct a detailed survey of the axisymmetric global eigenmodes computed around these base states. For sufficiently high Reynolds number, destabilisation occurs for an eigenmode on an ``arc branch’’ separated from other families of more stable eigenmodes. These arc branch modes, which we term ``flame-tip'' modes, are characterised by strong fluctuations near the flame tip and are independent of numerical domain truncation. A detailed, quantitative description of the linear feedback mechanism driving their destabilisation is provided by associating the frequency of the leading flame-tip mode with the Lagrangian advection time along the outer shear layer from the nozzle exit to the flame tip. A non-modal resolvent analysis demonstrates that the receptivity of the flame to forcing of the linear operator is largely consistent with simple resonance of the linear eigenmodes. Significant pseudo-resonance is not observed.

Upon assessing these linear results against nonlinear time-domain simulations, however, a more complex picture emerged. For small initial perturbations, the onset of sustained unsteadiness corresponds to the critical Reynolds number identified by linear analysis. Conversely, for sufficiently large perturbations, self-sustained oscillations occur even at Reynolds number values where the flame is linearly stable, revealing a substantial interval of hysteresis. Continuation analysis along this branch of unsteady solutions reveals an ordered sequence of state transitions in the subcritical regime. Along most of the unsteady branch, the unsteady flow settles into a limit-cycle state with a periodicity that does not match any linear eigenmodes of the base flow along the steady solution branch. However, as the Reynolds number approaches the critical value for linear instability of the steady state, the dynamics become enriched by an ordered sequence of increasingly high-dimensional features, including apparent quasi-periodicity and chaos. Interestingly, the frequency associated with the leading (stable) eigenmode of the base state becomes prominently represented in the power density spectrum during this process, suggesting a Neimark--Sacker bifurcation arising from two competitive modes. This dynamics is consistent with a Reulle--Takens--Newhouse scenario for the onset of chaos in the V-flame. Together, these findings shed new light on the nonlinear dynamical elements underpinning the observed V-flame behaviour. 

Building from the analysis of linear flame-tip mode, we carry out an analysis of the non-local interaction in the subcritical oscillation in the nonlinear regime. A Monte Carlo simulation of passive flow tracers is conducted, with tracers departing from the burner exit at various phase time and radial positions. Based on the hypothesis that advected vortex structures interact with the flame surface, we design criteria to test this hypothesis by selecting the conforming trajectories and estimating the nonlinear oscillation frequency based on the mean values of the convective time delay corresponding to each conforming trajectory. In agreement with our physical reasoning, the estimated frequency is found to be close to the reference nonlinear oscillation frequency when similarly stringent selection thresholds are employed simultaneously for the vorticity and heat release rate. The resulting trajectories intersect with the flame surface at phase instances where the flapping flame tip is significantly further upstream than in the steady base states.

Finally, we assess the capacity of linear methods to predict basic features of the V-flame dynamics, as is commonly attempted in the reduced-order modelling literature. As neither modal nor non-modal analysis of the steady base flow provide any hint about the observed subcritical limit-cycle oscillations, we attempt an eigenmode analysis of the time-averaged mean flow. This approach is hampered by the strong nonlinearity of the system, particularly visible in the reaction rate term. Both \textit{a priori} assessment and the computed mean flow eigenmodes reveal notable disparities in the mean flow eigenvalues and eigenmodes when compared to the reference Fourier modes associated with the nonlinear frequency of the limit-cycle fundamental. This result highlights known limitations of linear mean flow analysis, namely its failure to provide valid predictions for subcritical instabilities and in the presence of strong non-monochromatic interactions. Such issues may indeed be quite common in premixed laminar flame oscillations driven by flame--flow feedback due to the well-known interaction between convective velocity perturbations at a fundamental tone and the unsteady heat release rate at the first harmonic~\citep{lang1991harmonic,lieuwen2005nonlinear}. The present failure of mean flow analysis in the V-flame configuration may be contrasted with our successful earlier efforts in cylinder-stabilised flames \citep{wang2022global}, which featured supercritical bifurcation behaviour dominated by monochromatic hydrodynamic amplification mechanisms localised to the wake region without significant feedback from downstream heat release fluctuations.

\section*{Declaration of Interests} 
The authors report no conflict of interest.

\section*{Author ORCIDs}
\begin{itemize}
    \item C.-H. Wang: 0000-0002-4522-1320
    \item C. Douglas: 0000-0002-5968-3315
    \item Y. Guan: 0000-0003-4454-3333
    \item C.-X. Xu: 0000-0001-5292-8052
    \item L. Lesshafft: 0000-0002-2513-4553
\end{itemize}

\section*{Acknowledgements}
This study was inspired by the chapter ``Modal analysis of a V-shaped flame'' of C.-H. Wang's PhD thesis. For the original work in that chapter, C.-H. Wang and L. Lesshafft worked in collaboration with Grégoire Varillon and Wolfgang Polifke (TU Munich), who computed base states and conducted time-resolved simulation of V-flames with OpenFOAM. The authors would like to express their sincere gratitude for the scientific inputs from G. Varillon and W. Polifke in the original work that inspired the present study. C.-H. Wang was supported through a PhD scholarship from \'Ecole Polytechnique and by the Shuimu postdoc fellowship from Tsinghua University. C.-H. Wang acknowledges the China Postdoctoral Science Foundation (CPSF) and the National Natural Science Foundation of China (NSFC) for funding this work (Grant No. 2022TQ0181, 12388101). C. Douglas acknowledges funding for this project from the European Union's Horizon 2020 research and innovation program under the Marie Skłodowska--Curie Grant Agreement No. 899987. Y. Guan was supported by the PolyU Start-up Fund (Project No. P0043562) and the NSFC (Grant No. 52306166).

\bibliography{flame.bib}
\bibliographystyle{jfm}

\end{document}